\documentclass[
pra,reprint,nofootinbib,superscriptaddress,eqsecnum
]{revtex4-2}
\usepackage[T1]{fontenc}
\usepackage[english]{babel}
\usepackage{graphicx}
\usepackage{dsfont}
\usepackage{bm}
\usepackage{qcircuit}
\usepackage{physics}
\usepackage{mathtools}
\usepackage{amsmath}
\usepackage{amsfonts}
\usepackage{amssymb}
\usepackage{amsbsy}
\usepackage{hyperref}
\usepackage[capitalise]{cleveref}
\usepackage{wasysym}
\usepackage{physics}
\usepackage[hats]{./quantum_local}
\usepackage{bclogo}

\hypersetup{
    colorlinks=true,
    linkcolor=blue,
    urlcolor=blue,
    citecolor=blue,
}

\renewcommand{\l}{\op{\ell}}
\newcommand{\uGop}[1][]{\op{u}_{G #1}}
\newcommand{\vGop}[1][]{\op{v}_{G #1}}
\newcommand{\perx}{a}
\newcommand{\pery}{b}
\newcommand{\perr}{\frac{\pery}{\perx}}
\newcommand{\tfraa}{\fracpart{t}{\alpha}}
\newcommand{\tinta}{\closestint{t}{\alpha}}
\newcommand{\tfra}{\fracpart{t}{\pi/\alpha}}
\newcommand{\sfra}{\fracpart{s}{\alpha}}
\newcommand{\TUU}{\op{T}^U}
\newcommand{\TVV}{\op{T}^V}
\newcommand{\PUU}{\op{P}^U}
\newcommand{\PVV}{\op{P}^V}
\newcommand{\TUS}{\TUU_S}
\newcommand{\TVS}{\TVV_S}
\newcommand{\PUS}{\PUU_S}
\newcommand{\PVS}{\PVV_S}
\newcommand{\TUG}{\TUU_G}
\newcommand{\TVG}{\TVV_G}
\newcommand{\PUG}{\PUU_G}
\newcommand{\PVG}{\PVV_G}
\newcommand{\zakdom}{\mc{P}}
\newcommand{\gdom}{{\mc{P}_{G}}}
\newcommand{ \intgdom}{ \int_{\gdom} \hspace{-8pt} }
\newcommand{\GKP}{\mathrm{GKP}}
\newcommand{\gkpproj}{\op{\Pi}_{\rm GKP}}
\newcommand{\EC}{\mathrm{EC}}

\newcommand{\xfra}{\fracpart{x}{a}}
\newcommand{\yfra}{\fracpart{y}{2\pi/a}}
\newcommand{\xint}{\closestint{x}{a}}
\newcommand{\psquee}{\p_{\pery/\perx}}
\newcommand{\HCV}{\mc{H}_{\text{CV}}}
\newcommand{\util}{\tilde{u}}
\newcommand{\vtil}{\tilde{v}}
\newcommand{\synst}{\op{\Pi}\pqty{\util,\vtil}}
\newcommand{\eckraus}{\op K_\EC}
\newcommand{\ecchan}{\mathcal E_\EC}
\newcommand{\gkpec}{\eckraus \pqty{\util,\vtil}}

\begin{document}

\title{
The Zak transform: a framework for quantum computation with the Gottesman-Kitaev-Preskill code
}

\author{Giacomo Pantaleoni}

\email{g.pantaleoni@pm.me}

\affiliation{Centre for Quantum Computation \& Communication Technology, School of Science, RMIT University, Melbourne, VIC 3000, Australia}
\affiliation{Centre for Engineered Quantum Systems, School of Physics, University of Sydney, Sydney, NSW 2006, Australia}

\author{Ben Q. Baragiola}

\affiliation{Centre for Quantum Computation \& Communication Technology, School of Science, RMIT University, Melbourne, VIC 3000, Australia}
\affiliation{Center for Gravitational Physics and Quantum Information (CGPQI), Yukawa Institute for Theoretical Physics, Kyoto University, Kitashirakawa Oiwakecho, Sakyo-ku, Kyoto 606-8502, Japan}

\author{Nicolas C. Menicucci}

\affiliation{Centre for Quantum Computation \& Communication Technology, School of Science, RMIT University, Melbourne, VIC 3000, Australia}

\begin{abstract}
The Gottesman-Kitaev-Preskill (GKP) code encodes a qubit into a bosonic mode using periodic wavefunctions. This periodicity makes the GKP code a natural setting for the Zak transform, which is tailor-made to provide a simple description for periodic functions. We review the Zak transform and its connection to a Zak basis of states in Hilbert space, decompose the shift operators that underpin the stabilizers and the correctable errors, and we find that Zak transforms of the position wavefunction appear naturally in GKP error correction. We construct a new bosonic subsystem decomposition (SSD)---the modular variable SSD---by dividing a mode's Hilbert space, expressed in the Zak basis, into that of a virtual qubit and a virtual gauge mode. Tracing over the gauge mode gives a logical-qubit state, and preceding the trace with a particular logical-gauge interaction gives a different logical state---that associated to GKP error correction.
\end{abstract}

\maketitle

\section{Introduction}
\label{sec:intro}

The vast majority of schemes for quantum computation require discrete Hilbert spaces constructed from tensor products of $2$-dimensional systems---the discreteness is the basis of quantum error correction and fault tolerance as well as known quantum algorithms. This has led to the question of how infinite-dimensional continuous-variable systems, such as quantized electromagnetic fields~\cite{chuang_bosonic_1997,cochrane_macroscopically_1999}, superconducting circuits~\cite{wallraff_strong_2004,blais_circuit_2021}, or mechanical resonators~\cite{fluhmann_encoding_2019,leibfried_quantum_2003}, can be repurposed into behaving effectively as $2$-level systems (qubits). Clever methods to do so go beyond simply encoding qubits into continuous-variable systems; they also endow encodings with error-correcting properties, making them potential building blocks for a fault-tolerant quantum device. These encodings are collectively known as \emph{bosonic codes}.

The Gottesman-Kitaev-Preskill (GKP) code~\cite{gottesman_encoding_2001} stands out among bosonic codes due to its noise resilience~\cite{albert_performance_2018}, all-Gaussian implementation of logical Clifford gates~\cite{gottesman_encoding_2001} and magic-state production~\cite{baragiola_allgaussian_2019,matsuura_costreduced_2020}, seamless interface with continuous-variable cluster-state quantum computing~\cite{menicucci_faulttolerant_2014, walshe_streamlined_2021}, and rich set of mathematical properties~\cite{mensen_phasespace_2021, calcluth_efficient_2022}. The discretization mechanism at the basis of the GKP code is periodization in phase space: code-word wavefunctions in the computational and dual bases are periodic combs of Dirac-delta distributions in position and momentum, respectively. This introduces a redundancy that protects against small displacements in position and momentum, as the error correction procedure gives access to only small-displacement information while leaving the logical information untouched.

Pure GKP states, both ideal and approximate, are often represented as wavefunctions in a quadrature basis. An appealing alternative description is obtained by performing a Zak transform~\cite{zak_finite_1967,zak_dynamics_1968,zak_kqrepresentation_1972} of the position (or momentum) wavefunction~\cite{glancy_error_2006}. A Zak transform, also known as a Weyl-Brezin transform, takes a function of a single unbounded real variable to a function of two bounded real variables. This takes a wavefunction $\psi(x)$ with $x \in \mathbb{R}$ to a \emph{modular wavefunction} $\psi(u,v)$ with $u,v$ lying in a bounded patch of $\mathbb{R}^2$. The key property of modular wavefunctions is that they are particularly suited for compact representations not only of ideal, infinite-energy GKP states, but also their approximately periodic, finite-energy approximations.

The Zak transform has been used and re-discovered in mathematics (differential equations~\cite{gel1950expansion,brezin1970harmonic} and representation theory~\cite{weil1964certains}), physics~\cite{zak_finite_1967,zak_dynamics_1968,zak_kqrepresentation_1972}, and signal processing \cite{janssen_bargmann_1982,janssen_zak_1988,janssen_zak_1993}. In the early days of the GKP code, modular wavefunctions were used to describe approximate GKP states and GKP error correction~\cite{gottesman_encoding_2001,glancy_error_2006}, while more recently, they have found application in a wider range of GKP-related topics~\cite{terhal_encoding_2016,weigand_generating_2018,fluhmann_sequential_2018,le_doubly_2019,weigand_realizing_2020,matsuura_equivalence_2020}. Generalizations of the Zak transform also proved fruitful in the broader context of quantum information~\cite{ketterer_quantum_2014,ketterer_quantum_2016,albert_robust_2020,fabre_wigner_2020} and superconducting circuits~\cite{le_doubly_2019,le2020building}.

In this work, we delve further into connections between the Zak transform and the GKP code. The results are organized into two parts: the first part, \cref{sec:Zaktransform}, is devoted to the Zak transform, and the second part, comprising Sections~\ref{sec:gkpgeneral} and~\ref{sec:modvarSSD}, is devoted to the GKP code.

We begin in \cref{sec:zaktransform} by presenting the Zak transform and its relation to the ``Zak basis'' for representing quantum states in Hilbert space. We collect a number of useful facts about modular wavefunctions and elaborate on their natural $2\pi$-area domain induced by their periodicities. Modular arithmetic plays an important role throughout this work, especially so in the process of keeping track of the phasing rules that modular wavefunctions must obey. We provide simple formulas to keep track of these phases.

In \cref{sec:stretchedzak}, we show that the periodicity conditions can be relaxed by defining ``stretched Zak bases'' whose domain has arbitrary but finite area. We conclude the section by providing the link between \emph{modular variables}~\cite{aharonov_modular_1969,fabre_wigner_2020}---the operators that are diagonalized by the Zak-basis eigenstates---and the usual position and momentum operators of a bosonic mode.

In \cref{sec:gkpgeneral}, we introduce the GKP code in the modular-variable formalism and show how the Zak framework is inherently present both when considering the problem of evaluating the logical information of approximate GKP codewords and when performing GKP error correction. In \cref{sec:modvarSSD}, we develop a formalism specially tailored to deal with the problem of addressing the logical content of any continuous-variable state with respect to the GKP code. We use the stretched Zak basis to decompose the CV Hilbert space of a bosonic mode into that of a two-dimensional qubit and a infinite dimensional gauge subsystem. We refer to this change of basis as a subsystem decomposition (SSD)~\cite{raynal_encoding_2012,pantaleoni_modular_2020}. In SSDs, logical-qubit information is encoded in a virtual subsystem rather than in a subspace, which is the typical framework for quantum error correcting codes. We show that this decomposition gives a simple interpretation of the GKP error correction procedure, and also, somewhat surprisingly, that the logical content is identical to that the partitioned-position SSD~\cite{pantaleoni_modular_2020}.

\section{Definitions and properties}
\label{sec:Zaktransform}

\subsection{The Zak basis}
\label{sec:zaktransform}

The Zak transform maps a square-integrable function $\psi \in L^{2}(\reals)$ into a square-integrable, quasi-periodic function of two real variables with period $\perx$. Denoting with $\psi(x)$ the function $\psi$ evaluated at the point $x \in \reals$, its Zak transform $Z\psi$, evaluated at $u,v$ is~\cite{zak_finite_1967, zak_kqrepresentation_1972}
\begin{align}
\label{eq:zaktransform}
    (Z \psi)(u,v)
        =
            \sqrt{\frac{\perx}{2\pi}}
                \sum_{m \in \mathbb{Z}}
                    e^{ - i \perx m v} \psi\pqty{ u + \perx m }
                        \,.
\end{align}
For the moment, we let $u$ and $v$ take any real value. For the remainder of this work, we simply indicate $\psi(u,v)$ instead of $(Z \psi)(u,v)$, as there is no risk of confusion: $\psi(x)$ will always refer to a function in $L^2(\reals)$, and $\psi(u,v)$ will always refer to its Zak transform. Since we are mostly concerned with quantum states, we use bra-ket notation wherever possible and interpret square integrable functions as wavefunctions in the position representation, $\psi(x) = \braket{x}{\psi}[q]$, where $\ket{x}[q]$ is a position eigenstate such that
$
    \q\ket{x}[q]
        =
            x\ket{x}[q]
$,
$ x \in \reals $ (we reserve the symbol $\ket{x}[p]$ to momentum eigenstates). The position and momentum operators are $\op{q} \coloneqq \frac{1}{\sqrt{2}} (\op{a} + \op{a}^\dagger)$ and $\op{p} \coloneqq \frac {-i}{\sqrt{2}} (\op{a} - \op{a}^\dagger)$, respectively, in terms of creation and annihilation operators. We refer to $\psi(u,v)$ as a \emph{modular wavefunction} when it describes a pure quantum state. We note in passing that alternative phasings in the definition of the Zak transform, \cref{eq:zaktransform}, can be chosen, leading to different periodicity rules for modular wavefunctions---here, we use Zak's%
~\cite{zak_finite_1967,zak_dynamics_1968}.

We can construct a \emph{Zak ket} by applying the Zak transform to the position eigenstates, which gives a superposition of either position or momentum eigenstates
\begin{align}
\label{eq:zakvectordef}
    \ket{u,v}
        &=
            \sqrt{\frac{\perx}{2\pi}}
                \sum_{m \in \mathbb{Z}}
                    e^{ i \perx m v}
                        \qket{ u + \perx m }
                            \\
\label{eq:zakvectordefmom}
    &=
        \frac{1}{\sqrt\perx}
            e^{ -i u v }
                \sum_{m \in \mathbb{Z}}
                    e^{ - i \frac{2\pi}{\perx} m u }
                        \pket{ v + \tfrac{2\pi}{\perx} m }
                            \,,
\end{align}
where the normalization factor $(a/2\pi)^{-1/2}$ ensures orthonormality in the Dirac-comb sense
\begin{equation}
    \bra{u,v} \ket{u',v'}
        =
            \sum_m\delta(u-u' + \perx m)
                \sum_n \delta(v-v'+ \tfrac{2\pi}{\perx}n)
                    \,.
\end{equation}
The reader interested in verifying \cref{eq:zakvectordefmom} will find Poisson's formula useful.

\begin{figure}[t]
    \includegraphics[width=0.45\textwidth]{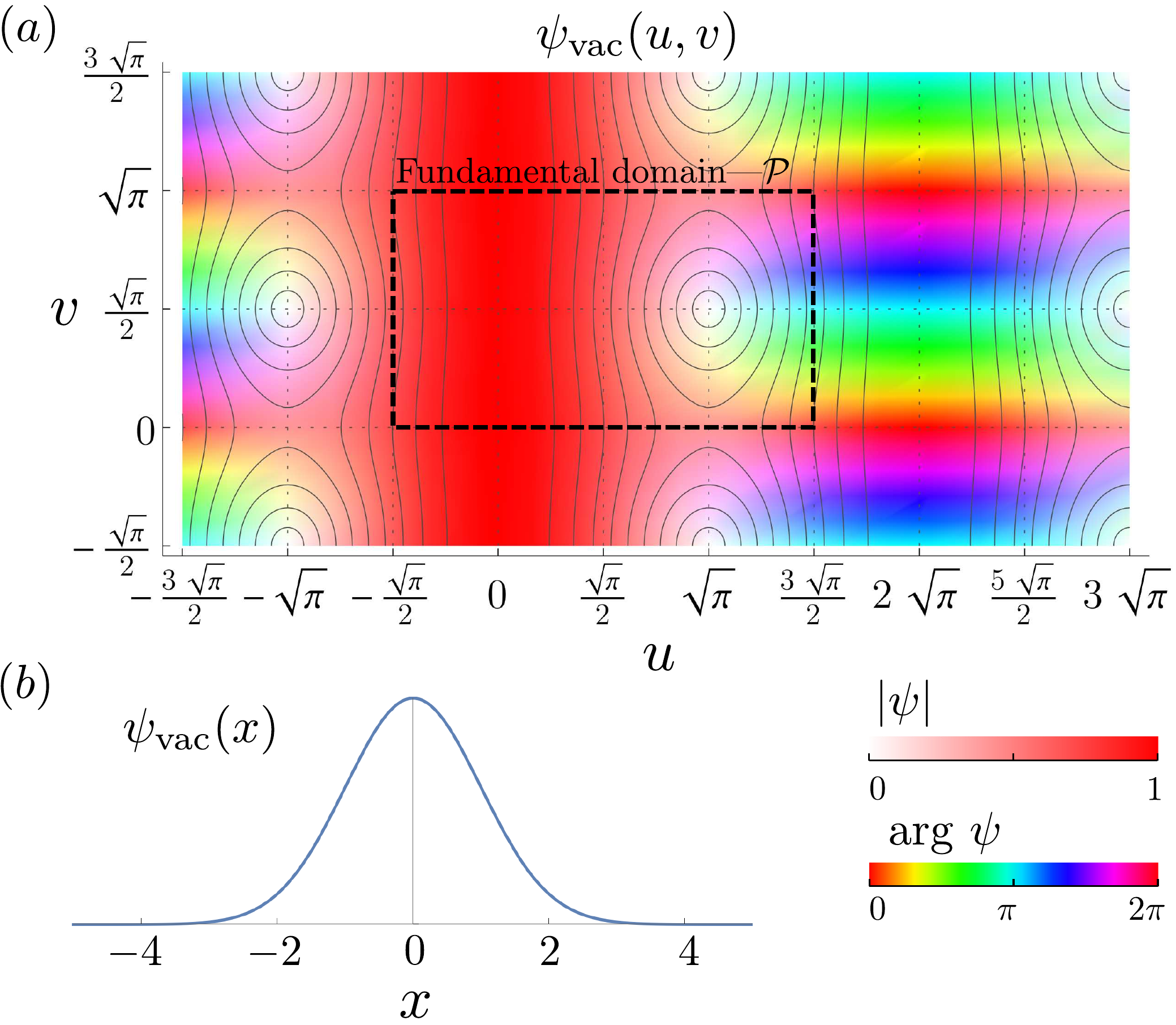}
    \caption{%
        $(a)$ Modular wavefunction $\psi_\text{vac}(u,v)$ and $(b)$ position wave function
        $
            \psi_\text{vac}(x)
                \propto
                    \exp(-x^2/2)
        $
        for the vacuum state of the harmonic oscillator. The modular
        wavefunction is given for a Zak transform with period $\perx=2\rpi$. It
        is periodic in the vertical direction, and periodic modulo a phase
        (quasi-periodic) in the horizontal direction. Because of the
        quasi-periodicity, the values of $\psi_\text{vac}(u,v)$ outside
        $\zakdom$ are redundant and it is sufficient to restrict ourselves to a
        fundamental domain $\zakdom$ whose center can be freely chosen. The
        choice of periodicity and centering here is convenient for representing
        states of the square GKP code, whose code words have a $2\rpi$
        periodicity and support only on integer multiples of $\rpi$.
    }
    \label{fig:zakvacuuma}
\end{figure}

In analogy with the position and momentum wavefunctions in their respective bases, we refer to $\psi(u,v)$ in~\cref{eq:zaktransform} as the wavefunction in the Zak representation (or, more simply, the modular wavefunction) by identifying it with the inner product
\begin{align}
    \psi(u,v)
        \coloneqq
            \bra{u,v} \ket{\psi}
                \,,
\end{align}
One may verify that this expression is consistent by taking the adjoint of~\cref{eq:zakvectordef} and using the inner product of $L^{2}(\reals)$. The dual vector $\bra{u,v}$ is thus the linear functional that gives a wavefunction in Zak representation evaluated at $u,v$, in the same sense that $\bra{x}[q]$ and $\bra{x}[p]$ are the linear functionals that give the wavefunction in the position and momentum bases evaluated at the point $x$. An example modular wavefunction, that of the vacuum state, is shown in \cref{fig:zakvacuuma}.

\begin{figure}[t]
    \includegraphics[width=0.3\textwidth]{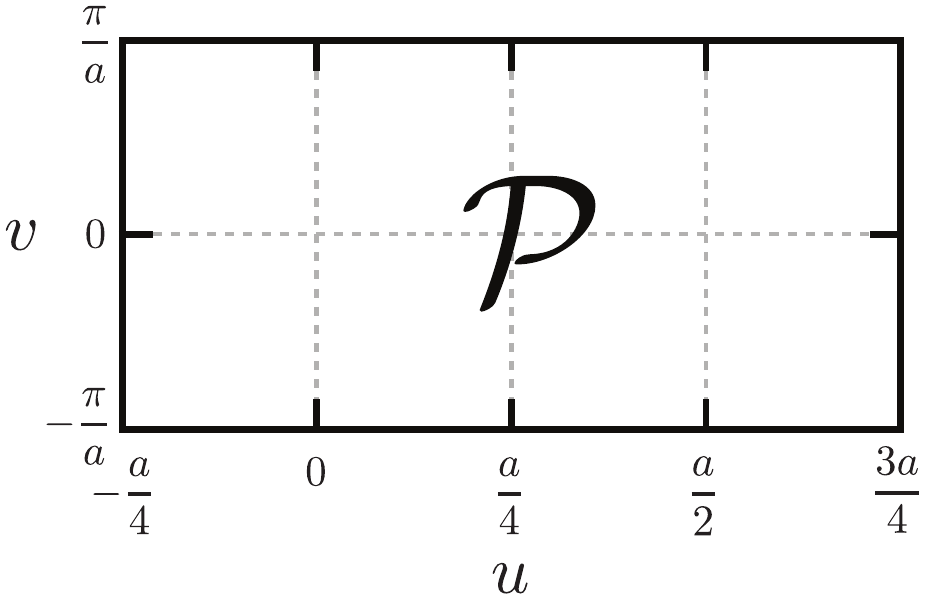}
    \caption{%
        Centered fundamental Zak domain $\zakdom$ with width $a$ and height
        $2\pi/a$.
    }
\label{fig:centeringa}
\end{figure}

Two important properties of the vectors in \cref{eq:zakvectordef} are quasi-periodicity in the first variable and periodicity in the second,
\begin{subequations}
\label{eq:Zakperiodicities}
    \begin{align}
    \label{eq:zakketperiodu}
        \ket{u+\perx,v}
            &=
                e^{-i\perx v} \ket{u,v}
                    \\
    \label{eq:zakketperiodv}
        \ket{u,v + 2\pi / \perx}
                &=
                    \ket{u,v}
                        \,.
    \end{align}
\end{subequations}
which are inherited by modular wavefunctions $\psi(u,v)$,
\begin{align}
    \label{eq:zakperiodv}
        \psi\pqty{u + \perx,v}
            &=
                e^{i\perx v} \psi(u,v)
                    \\
    \label{eq:zakperiodu}
        \psi(u,v + 2\pi / \perx)
            &=
                \psi(u,v)
                    \,.
\end{align}
An important consequence of these properties is that the states $\ket{u,v}$ form an overcomplete basis when there are no restrictions on the domain of $u$ and $v$.  The standard prescription to construct an orthonormal Zak basis is to restrict the domain of $u$ and $v$ to a rectangle of area $2\pi$, which we often refer to as a ``Zak patch'', whose sides are given by the periods $\perx$ and $2\pi/\perx$. This is equivalent to restricting the domain of modular wavefunctions to a torus. One is free to choose the centering of the Zak patch; we choose
\begin{equation}
\label{eq:zakdom}
    \zakdom
        =
            \Big[
                -\frac{\perx}{4}, \frac{3\perx}{4}
            \Big)
                \times
                    \Big[
                        -\frac{\pi}{\perx}, \frac{\pi}{\perx}
                    \Big)
\end{equation}
so that the two points $(u,v)=(0,0)$ and $(u,v) =(\perx,0)$, which will be important for GKP states, lie within the patch and not on its boundaries. The fundamental Zak patch is shown in \cref{fig:centeringa}.

The states $\ket{u,v}$ with $u,v\in\zakdom$, span the Hilbert space of a bosonic mode, and one may write the completeness as~(see Ref.~\cite{ketterer_quantum_2016} for a brief, focused discussion and Ref.~\cite{bacry_proof_1975} for a rigorous proof)
\begin{align}
\label{eq:completeness}
    \int_{\zakdom} du\, dv\,
        \ket{u,v}\bra{u,v}
            =
                \int_{\reals} dx
                    \ket{x}[q]\bra{x}[q]
                        =
                            \op \id
                                \,.
\end{align}
The Zak transform is then interpreted as an isometry from the Hilbert space of complex-valued, square-integrable functions on the real line to the Hilbert space of complex-valued, square-integrable functions of two real variables in the Zak domain $\zakdom$, $L^{2}(\reals) \rightarrow L^{2}(\zakdom)$~\cite{janssen_zak_1988,janssen_zak_1993,grochenig2001foundationstimefrequency}. With the restriction on the domain to $\zakdom$, the orthonormality condition becomes
\begin{align}
    \label{eq:orthonormalizy}
        \bra{u,v}\ket{u',v'}
            =
                \delta^{(\perx)}(u-u')
                    \delta^{(2\pi/\perx)}(v-v')
    \,,
\end{align}
where $\delta^{(\perx)}(u)$ and $\delta^{(2\pi/\perx)}(v)$ are Dirac delta-distributions in the horizontal and vertical intervals of the Zak patch respectively. The \emph{Zak basis} is then
\begin{align}
\label{eq:zakbasis}
    \mc{B}_{Z}
        =
            \{\ket{u,v}\
                |\ u,v \in \zakdom
            \}
                \,.
\end{align}
Unless otherwise stated, from now on, when we use the symbols $u$ and $v$, it is understood that $u,v \in \zakdom$.

Zak states that lie outside of $\zakdom$ are phased versions of the Zak-basis states within $\zakdom$. The periodicity and quasi-periodicity conditions give the recipe to find this phase. First, recall that a real number $x \in \mathbb{R}$ can be written as a quotient and remainder with respect to a positive real number $\delta$. Including a centering $\mu$, $x$ decomposes as
\begin{align}
\label{eq:intfrac}
    x
        =
           \fracpart{x}{T}^\mu
               +
                   \closestint{x}{T}^\mu
                       \,,
\end{align}
where $\fracpart{x}{T}^\mu \in [-\mu,T-\mu)$ is the centered fractional part of $x$ and $\closestint{x}{T}^\mu = x - \fracpart{x}{T}^\mu$ is the centered closest integer multiple of $T$ to $x$. We will omit the centering superscript whenever the centering does not matter. Consider now a Zak state $\ket{x,y}$, where $x,y \in \mathbb{R}$ are not limited to $\zakdom$. By decomposing $x$ and $y$ and using \cref{eq:zakketperiodu,eq:zakketperiodv}, one finds that
\begin{align}
\label{eq:zakcentering0}
    \ket{x,y}
        &=
            e^{-i \xint^{\perx/4} \yfra^{\pi/\perx} }
                \ket{\xfra^{\perx/4}, \yfra^{\pi/\perx}}
                    \,,
\end{align}
with the state on the right-hand side being a bona fide element of the orthonormal Zak basis within $\zakdom$, \cref{eq:zakbasis}. An example that appears often is a Zak basis state $\ket{u,v}$ with $u,v \in \zakdom$ that has undergone shifts by unrestricted values $s$ and $t$,
\begin{align}
\label{eq:zakcentering}
    \ket{ u + s , v + t }
        =
            e^{
                    -i \closestint{ u + s }{a} \fracpart{ v + t }{2\pi/a}
                }
                    \ket{
                        \fracpart{ u + s }{a} ,
                            \fracpart{ v + t }{2\pi/a}
                        }
                            \,.
\end{align}

We discuss the effect of quadrature shifts more in depth, as well as their Zak space versions. The Weyl-Heisenberg shift operators
\begin{align}
\label{eq:WHshift}
    \Z\pqty{t}
        &\coloneqq
            e^{i \q t}
                \,, \\*
    \X\pqty{t}
        &\coloneqq
            e^{-i \p t}
                \,,
\end{align}
respectively phase and shift position eigenstates, $\Z\pqty{t} \qket{x} = e^{i x t} \qket{x}$
and
$
    \X\pqty{t} \qket{x}
        =
        \qket{x + t}
$,
with complementary actions on momentum eigenstates.
In the Zak basis, their actions are
\begin{align}
\label{eq:zuv}
    \Z\pqty{t} \ket{u,v}
        &=
            e^{iut} \ket{ u, v + t }
                \,, \\
\label{eq:xuv}
    \X\pqty{t} \ket{u,v}
        &=
            \ket{ u + t , v}
                \,.
\end{align}
We can see that, once a fundamental comb state is defined,
$
\ket{0,0}
    \coloneqq
        \sum_{m \in \mathbb{Z}}
            \qket{ \perx m }
$,
an equivalent definition of the Zak transform can be given. We can do so by defining a Zak vector as a displaced $\ket{0,0}$ state,
\begin{align}
\label{eq:Zakstatedef}
    \ket{u,v}
        \coloneqq
            \op{X}(u) \op{Z}(v) \ket{0,0}
                \,,
\end{align}
and $\psi(u,v) = \bra{u,v}\ket{\psi}$ as the Zak transform of $\ket \psi$.
This alternative definition highlights the fact that we have taken the convention where momentum shifts are performed first. Different ordering and phasings of the displacements would give different Zak vectors. There are two more conventions that appear to be just as natural or useful as the one above. The second choice is the one where the shifts are taken in the opposite order $\ket{u,v}_\text{op} \coloneqq \op{Z}(u)\op{X}(v)\ket{0,0}$, and in the third one they are performed symmetrically $\ket{u,v}_\text{sym} \coloneqq e^{i(v \op q - u \op p)} \ket{0,0}$. These Zak states differ from those defined in \cref{eq:Zakstatedef} by phases, $\ip{u,v}{u,v}[\text{op}] = e^{-iuv}$ and $\ip{u,v}{u,v}[\text{sym}] = e^{-i \frac{uv}{2}}$. One could proceed using any of these conventions; we use \cref{eq:Zakstatedef}.

\begin{figure*}[t]
    \includegraphics[width=0.9\textwidth]{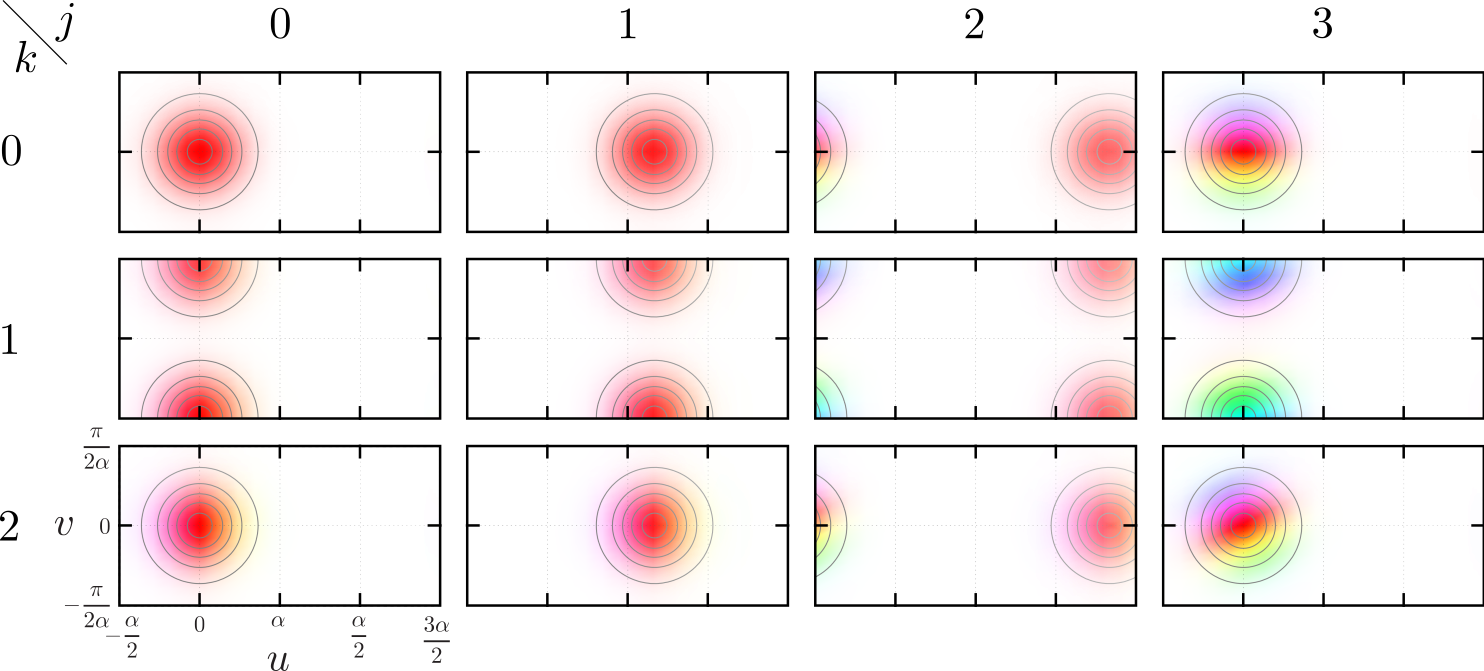}
    \caption{%
        Effect of combined position and momentum displacements,
$ \X \big( j \frac{\rpi }{3}  \big) \Z \big( k\frac{\rpi }{2} \big) $ for
integer $j$ and $k$, on the modular wavefunction of an approximate GKP state.
Dimensions are specified in the (2,0) plot, and are the same in every plot.
We consider an approximation to a GKP-$\ket{0}$ obtained from (here we are
referring to the position representation) a comb of Gaussian functions with
spacing $2\rpi$. Each Gaussian has variance $\Delta^2$, and the comb is
multiplied by an enveloping Gaussian with variance $\Delta^{-2}$. These
variances make the vertical and horizontal spread of the modular wavefunction
the same. The actual value variance is chosen for visual clarity. Each subplot
shows a modular wavefunction plotted in a fundamental domain $\zakdom$,
\cref{eq:zakdom}, with $a = 2\sqrt{\pi}$ (equivalently, $\alpha = \rpi$ in
\cref{fig:centeringalpha}), and the color legend is the same as in
\cref{fig:zakvacuuma}. The columns in the array of subplots correspond to values
of $j$ and the rows to values of $k$. The acquired phases can be understood with
the help of the relations in \cref{eq:xuv,eq:zuv}. Notice that the
difference in relative phasing across the support of the state encodes
displacements by a complete patch length or width. For instance, compare the
relative phasing across the states in the four corners, corresponding to
$(j,k) \in \{(0,0), (0,3), (3,0), (3,3)\}$. These states would be identical
without this difference in phasing. It is solely this phasing that encodes the
relative displacement between these four states. Finally, note that in the ideal
case of a GKP-$\ket 0$ state (Dirac delta at the origin of $\mathcal P$),
displacements by a full patch have no effect, and this is reflected in the fact
that the origin of all four of these states remains unchanged in phase.
    }
    \label{fig:shifts_agkp}
\end{figure*}

We introduce operators that produce phases and shifts on Zak eigenstates, analogue to the action of Weyl-Heisenberg operators on position eigenstates.
We define \emph{modular phase operators}
\begin{subequations}
\label{eq:modphaseops}
    \begin{align}
    \label{eq:uvdisp1}
        \PUU(t)
            \ket{u,v} &
                \coloneqq
                    e^{i u t}\ket{u,v}
                        \,, \\
    \label{eq:uvdisp2}
        \PVV(t)
            \ket{u,v} &
                \coloneqq
                    e^{i v t}\ket{u,v}
                        \,,
    \end{align}
\end{subequations}
for $u,v\in\zakdom$ and $t \in \reals$.
The modular phase operators are generated by the \emph{modular variables} $\op u$ and $\op v$, $ e^{i \op{u} t} = \PUU(t) $ and $ e^{i \op{v} t} = \PVV(t) $. The \emph{modular shift operators} are defined as
\begin{subequations}
\label{eq:modshiftops}
    \begin{align}
    \label{eq:uvdisp3}
        \TUU(t)
            \ket{u,v} &
                \coloneqq
                    \ket{u+t,v}
                        \,, \\
    \label{eq:uvdisp4}
        \TVV(t)
            \ket{u,v} &
                \coloneqq
                    \ket{u,v+t}
                        \,,
    \end{align}
\end{subequations}
for $u,v\in\zakdom$ and $t \in \reals$. Here, an asymmetry becomes evident. Scaled integer position $a \op{m}$ generates a modular momentum translation, $ e^{i \perx \op m t} = \TVV(t) $, but scaled integer momentum $\frac{2 \pi}{a} \op{n}$ generates a composite action, $ e^{- i \frac{2\pi}{\perx} \op{n} t} = \PVV(t) \TUU(t) $. It is not possible to generate translations in modular position by exponentiating only one of the fundamental operators in \crefrange{eq:aha1}{eq:aha4}. This is because, with our choice of convention for the Zak transform and Zak states (introduced by Zak~\cite{zak_dynamics_1968,zak_kqrepresentation_1972}), modular position and modular momentum are not on equal footing, at least when interpreting them in terms of Aharonov's integer and modular operators. For this reason, the quadrature shift operators decompose as
\begin{subequations}
\label{eq:modularshifts}
    \begin{align}
    \label{eq:zmodular}
        \Z(t)
            &=
                \PUU(t) \TVV(t)
                    \,, \\
    \label{eq:xmodular}
        \X(t)
            &=
                \TUU(t)
    \end{align}
\end{subequations}
which are simply \cref{eq:zuv,eq:xuv} in a basis-independent form.%
\footnote{%
    It is possible to work with more symmetrical relations using the Zak kets,
    $
        \ket{u,v}_\text{sym}
            =
                e^{\frac{iuv}{2}}\ket{u,v}
    $.
    We then get a more symmetrical decomposition of the Weyl-Heisenberg
    operators,
    $
        \X(t)
            =
                \TUU_\text{sym}(t)
                    \PVV_\text{sym}(-t/2)
    $
        and
    $
        \Z(t) = \TVV_\text{sym}(t)  \PUU_\text{sym}(t/2)
    $,
    with the modular shift and phase operators acting on symmetric Zak states
    as in \crefrange{eq:uvdisp1}{eq:uvdisp4}. Exponentiating the scaled integer
    operators generates a combination of primed modular phases and shift,
    $
        e^{i \perx \op m t}
            =
                \TVV_\text{sym}(t)
                    \PUU_\text{sym}(-t/2)
    $
    and
    $
        e^{i \frac{2\pi}{\perx}\op n t}
            =
                \TUU_\text{sym}(t)\PVV_\text{sym}(t/2)
    $.
    We do not adopt this convention throughout this work.
}
We illustrate their action on modular wavefunctions in \cref{fig:shifts_agkp}.

The modular phase and modular shift operators obey similar commutation relations to that of the Weyl-Heisenberg operators,
$
    \Z(s)\X(t)
        =
            e^{ist}
                \X(t)\Z(s)
$.
That is, the only non-zero commutators between pairs of modular operators come from shifts and phases of the same modular variable:
$
    \PUU(s){\TUU(t)}
        =
            e^{ist}
                \TUU(t){\PUU(s)}
$
and
$
    \PVV(s){\TVV(t)}
            =
                e^{ist}
                    \TVV(t){\PVV(s)}
$.

The conditions in \cref{eq:zakketperiodu,eq:zakketperiodv} imply periodicity on the modular shift operators in the following sense:
\begin{align}
\label{eq:zakperopU}
    \TUU(\perx)
        &=
            \PVV(-\perx)
                \,, \\
\label{eq:zakperopV}
    \TVV(2\pi/\perx)
        &=
            \id
                 \,.
\end{align}
With these relations, the modular shift operators can be rewritten using modular arithmetic as
\begin{align}
\label{eq:zakperopUmod}
    \TUU (t)
        &=
            \PVV ( -\closestint{t}{\perx} )
                \TUU ( \fracpart{t}{\perx} )
                    \,, \\
\label{eq:zakperopVmod}
    \TVV (t)
        &=
            \TVV ( \fracpart{t}{2\pi/\perx} )
                \,.
\end{align}
According to \cref{eq:zakperopUmod}, translations that wrap around the $u$-domain are accompanied by a phase. Translations on the $v$-domain do not exhibit this behavior, as shown in \cref{eq:zakperopVmod}. Note that these relations give us an alternative approach to obtaining \cref{eq:zakcentering0}.

\subsubsection{Relation to modular variables}

The modular variables $\op u$ and $\op v$ we encountered are an important concept in physics. They may be interpreted as non-local analogues of the position and momentum of a quantum particle~\cite{aharonov_modular_1969}. They are useful quantities in solid state physics, where useful dynamical variables are not necessarily local~\cite{zak_dynamics_1968,zak_kqrepresentation_1972} (in that context, they are known as quasi-position and quasi-momentum) and they have been used for interpreting quantum mechanical effects with no classical analogous, such as the Aharonov-Bohm effect. Here, we give explicit relations between the Zak kets, the operators diagonalizing them, and modular variables as known in the physics literature~\cite{aharonov_modular_1969}.

The operators $\op{u}$ and $\op{v}$ are simultaneously diagonalized by the Zak basis
\begin{align}
\op{u}\ket{u,v}
    &=
        u \ket{u,v}
            \,, \\
\op{v}\ket{u,v}
    &=
        v \ket{u,v}
            \,.
\end{align}
For all $u,v \in \zakdom$, they commute,
\begin{equation}
\label{eq:zakcommutator}
        [\op u, \op v] = 0
            \,,
\end{equation}
and are, in fact, modular position and momentum operators, respectively.
This can be shown by using \cref{eq:intfrac} to decompose the position and momentum operators following the prescription (with the exception of the centering) in Aharonov \emph{et al.}~\cite{aharonov_modular_1969},
\begin{align}
\label{eq:Aharonov}
    \q
        &=
            \op{u} + \perx \op{m}
                \,, \\*
        \p
            &=
                \op{v} + \frac{2\pi}{\perx} \op{n}
                    \,.
\end{align}
The modular position and momentum operators, $\op{u}$ and $\op{v}$, and integer position and momentum operators, $\op{m}$ and $\op{n}$, are given by%
\begin{subequations} \label{eq:modoperators}
\begin{align}
\label{eq:aha1}
    \op{u}
        &\coloneqq
            \fracpart{\q}{\perx}^{\perx/4}
                \,, \\
\label{eq:aha2}
    \op{v}
        &\coloneqq
            \fracpart{\p}{2\pi/\perx}^{\pi/\perx}
                \,, \\
\label{eq:aha3}
    \op{m}
        &\coloneqq
            \frac{\closestint{\q}{\perx}^{\perx/4}}{\perx}
            \,, \\
\label{eq:aha4}
    \op{n}
        &\coloneqq
            \frac{\closestint{\p}{2\pi/\perx}^{\pi/\perx}}{2\pi/\perx}
                \,.
\end{align}
\end{subequations}
The operators $\op{m}$ and $\op{n}$ have integer spectra, while the spectra of their counterparts, $\closestint{\q}{\perx}^{\perx/4}$ and $\closestint{\p}{2\pi/\perx}^{\pi/\perx}$, are integer multiples of the bin size.
The positive integer $\perx$ is interpreted as a ``bin size'' in position, with a complementary bin size in momentum of $2\pi/\perx$.

These formulas are useful to keep in mind. For example, evaluating the modular phase operators on unrestricted $\ket{x,y}$ state is potentially ambiguous, but we can do so using their expressions in terms of $\q$ and $\p$, \cref{eq:aha1,eq:aha2}, which give
\begin{align}
    \label{eq:modphasesunrestricted}
        \op{P}_{U}(t) \ket{x,y}
            &=
                e^{i t \xfra } \ket{x,y}
                    \,, \\
        \op{P}_{V}(t) \ket{x,y}
            &=
                e^{i t \yfra } \ket{x,y}
                    \,.
\end{align}

Although we do not make use of it in this work, a useful representation of the modular and integer operators, \crefrange{eq:aha1}{eq:aha4}, in differential form is given by the formal relations \begin{align}
    \bra{u,v} \op{u}
        &= u\,
            \bra{u,v}
                \,, \\
    \bra{u,v} \op{v}
        &=
            v \, \bra{u,v}
                \,, \\
    \bra{u,v} \perx \op m
        &=
            i \pdv{}{v} \bra{u,v}
                \,, \\
    \bra{u,v} \frac{2\pi}{\perx} \op n
        &=
            - \left(
                i \pdv{}{u} + v
            \right)
                \bra{u,v}
                    \,.
\end{align}

\subsection{The stretched Zak basis}
\label{sec:stretchedzak}

In \cref{sec:zaktransform}, we introduced and discussed the Zak transform as a map from functions $\psi$ on the real line to quasi-periodic functions of two variables, $Z \psi$. The Zak transform $Z$ is parametrized by a periodicity $\perx$, and the domain of $Z\psi$ can be restricted to a patch of the real plane with fundamental fixed area $2\pi$~\cite{bacry_proof_1975, janssen_bargmann_1982, barbieri_zak_2015}, rather than the whole plane itself, since $Z\psi$ is periodic.

In this section, we modify the definition of the Zak transform $Z$ to allow two independent periodicities, $\perx$ and $2\pi/\pery$, thereby stretching the area of the fundamental Zak domain to $2 \pi (a/b)$. We collect several formulas of practical interest that will be useful in \cref{sec:uvdec}, when presenting a subsystem decomposition of the CV Hilbert space into two subsystems. One of the two subsystem will be $2$-dimensional, while the other one will be a Zak-like domain with non-standard periodicity. Finally, we discuss the physical implications of a Zak domain with non-standard periodicity and show how the relationship between modular operators and quadrature operators is different from the standard case.

We define a \emph{stretched Zak basis} by introducing an additional periodicity parameter, $\pery>0$. Namely,
\newcommand{\stre}{S}
\begin{align}
\label{eq:stretchedbasis1}
    \ket{u,v}[\stre]
        &=
            \sqrt{\frac{b}{2\pi}}
                \sum_{m \in \mathbb{Z}}
                    e^{ i \pery m v}
                        \ket{ u + \perx m }[q]
                            \\
        &=
            \sqrt{\frac{1}{\pery}}
                e^{-i \frac{\pery}{\perx} u v}
                    \sum_{m \in \mathbb{Z}}
                        e^{
                            - i \frac{2\pi}{\perx} m u
                        }
                            \ket{
                                v + \frac{2\pi}{\pery m} }[p]
                                    \,,
    \end{align}
with subscript $S$ indicating stretched-basis vectors. The stretched Zak basis states are related to the standard Zak basis states, \cref{eq:zakvectordef}, as
\begin{align}
\label{eq:stretchedbasis2}
    \ket{u,v}[\stre]
        &=
            \sqrt{\perr}
                \ket{u, \perr v}
                    \,,
    \end{align}
similar in form to a squeezing transformation acting on the second argument, which we discuss in more detail later.

The stretched Zak basis exhibits a modified quasi-periodicity compared to \cref{eq:Zakperiodicities}. A shift in the first modular variable $u$ by the full period $\perx$ gives an acquired phase that depends on the second period $\pery$. Meanwhile, the periodicity in $v$ is identical to that of the standard Zak basis. Together,
\begin{subequations} \label{eq:abperiods}
\begin{align}
\label{eq:abperiod1}
    \ket{u + \perx,v}[S]
        &=
            e^{- i \pery v}
                \ket{u,v}[S]
                    \\
\label{eq:abperiod2}
    \ket{u,v + 2 \pi / \pery}[S]
        &=
            \ket{u,v}[S]
                \,.
\end{align}
\end{subequations}
The stretched Zak domain $\zakdom_{S}$ is a rectangle with sides given by the two periodicities $\perx$ and $2\pi/\pery$,
\begin{align}
    \zakdom_{S}
        =
            \Big[
                -\frac{1}{4} \perx ,
                 \frac{3}{4} \perx
            \Big)
                \times
             \left[
                -\frac{\pi}{\pery} ,
                 \frac{\pi}{\pery}
            \right)
                \,,
\label{eq:patchab}
\end{align}
where the centering is chosen for later convenience. The $2\pi (\perx/\pery)$ area of the stretched Zak patch depends explicitly on the periodicities, in contrast to standard Zak patch $\zakdom$, where $\perx = \pery$.

We define the stretched Zak basis in analogy to the standard one,
\begin{align}
\mc{B}^{S}_{Z}
    =
        \{\ket{u,v}[S]\
            |\ u,v \in \zakdom_S  \}
                \,,
\end{align}
With simple changes of variables, it can be shown that the basis states in $\mc{B}^{S}_{Z}$ are $\delta$-normalized
\begin{align}
\label{eq:orthonorm_stretched}
    \bra{u,v}\ket{u',v'}[S][S]
        =
            \delta^{(\perx)}(u-u')
                \delta^{(2\pi/\pery)}(v-v')
                    \,,
\end{align}
and complete
\begin{align}
    \int_{ \zakdom_{S} } du\, dv\,
        \ketbra{u,v}{u,v}[S][S]
            =
                \int_{ \zakdom } du\, dv\,
                    \ketbra{u,v}{u,v}
            =
                \id
                    \,.
\end{align}

The action of modular phase and translation operators in the stretched picture is the same as in \crefrange{eq:uvdisp1}{eq:uvdisp4}:
\begin{align}
\label{eq:uvstretchedopsPUG}
    \PUS( t ) \ket{u,v}[S]
        &=
            e^{i u t}\ket{u,v}[S]
                \,, \\
\label{eq:uvstretchedopsPVG}
    \PVS( t ) \ket{u,v}[S]
        &=
            e^{i v t}\ket{u,v}[S]
                \,, \\
\label{eq:uvstretchedopsRecU}
    \TUS( t ) \ket{u,v}[S]
        &=
            \ket{u + t,v}[S]
                \,, \\
\label{eq:uvstretchedopsRecV}
    \TVS( t ) \ket{u,v}[S]
        &=
            \ket{u,v+t}[S]
                \,,
\end{align}
but the periodicity of the stretched translation operator is different. Namely,
\begin{align}
    \label{eq:stretchedopperiods}
    \TUS( \perx )
        &=
            \PVS( -\pery )
                \\
        \TVS( \pi/\pery )
            &=
                \op \id_G
    \,,
\end{align}
which can be seen by using \cref{eq:abperiod1,eq:abperiod2}.

Just like the standard Zak basis, the stretched Zak basis has a set of operators associated with modular and integer quadratures. Stretched modular and integer position are the same as for the standard Zak basis, \cref{eq:aha1,eq:aha2},
\begin{align}
\label{eq:stretched_aha1}
    \op{u}_S &\coloneqq
        \fracpart{\q}{\perx}^{\perx/4}
            \,, \\
\label{eq:stretched_aha3}
    \op{m}_S
        &\coloneqq
            \frac{\closestint{\q}{\perx}^{\perx/4}}{\perx}
    \,.
\end{align}
As hinted by \cref{eq:stretchedbasis2}, the other set of modular and integer operators are associated with a \emph{squeezed} momentum operator. The squeezing operator is $\op{S}(\zeta) \coloneqq e^{\frac{i}{2} (\ln \zeta) (\op q \op p + \op p \op q)/2}$, so that the squeezed position and momentum operators are
\begin{align}
    \op{q}_\zeta
        &\coloneqq
            \op{S}^\dagger(\zeta) \op{q} \op{S}(\zeta)
                =
                    \zeta \op q
                        \,, \\
    \op{p}_\zeta
        &\coloneqq
            \op{S}^\dagger(\zeta) \op{p} \op{S}(\zeta)
                =
                    \zeta^{-1} \op p
                        \,.
\end{align}
The stretched modular and integer operators are those arising from momentum squeezed by $\zeta = b/a$,
\begin{align}
\label{eq:stretched_aha2}
    \op{v}_S &\coloneqq
        \fracpart{ \psquee }{2\pi/\pery}^{\pi/\pery}
            \,, \\
\label{eq:stretched_aha4}
    \op{n}_S &\coloneqq
        \frac{\closestint{ \psquee }{2\pi/\pery}^{\pi/\pery}}{2\pi/\pery}
            \,.
\end{align}
The above expressions for $\op{u}_S$ and $\op{v}_S$ along with \cref{eq:stretchedbasis2} give the expected eigenvalue equations,
\begin{align}
    \op{u}_S \ket{u,v}[S]
        &=
            u \ket{u,v}[S]
                \\
    \op{v}_S\ket{u,v}[S]
        &=
            v \ket{u,v}[S]
                \,.
\end{align}
The relationship between Aharonov's modular momentum $\op v$---defined as a fractional part with respect to $2\pi/\perx$---and the stretched modular momentum $\op{v}_S$ is given by
\begin{equation}
    \op{v}_S
        =
            \fracpart{ \psquee }{
                2\pi / \pery
            }^{\pi/\pery}
        =
            \frac{a}{b}
                \fracpart{ \op{p} }{
                    2\pi / \perx
                }^{\pi/\perx}
        =
            \frac{a}{b} \op{v}.
\end{equation}
This relation is the counterpart to the one the states in \cref{eq:stretchedbasis2}.
Once the modular-position period $\perx$ is chosen, $\op v_S$ has two equivalent interpretations: (1)~$\op v_S$ is modular momentum rescaled with respect to $2\pi/\perx$, or (2)~$\op v_S$ is modular squeezed momentum with respect to $2\pi/\pery$. In the first case, the rescaling factor is $a/b$, while in the second case, this same quantity is interpreted as a squeezing factor. We learn that, should we want to work with stretched modular variables, we may always relate them to Aharonov's with a squeezing operation.

That the operators $\op{u}_S$ and $\op{v}_S$ commute is not apparent since we have changed the modularity in the latter operator (it is no longer $2\pi/a$). In fact, the vanishing commutator of $\op u$ and $\op v$ defined with fixed modularities ($a$ in one quadrature and $\frac{2 \pi}{a}$ in the other) extends to arbitrary modularities when the quadratures are squeezed appropriately. To find a general expression, we multiply the commutator in \cref{eq:zakcommutator} by two positive real numbers $c_1$ and $c_2$ and make use of the modular-arithmetic expression
$
    c\fracpart{x}{T}^{\mu}
        =
            \fracpart{c x}{c T}^{c \mu}
$ to obtain vanishing commutation relations in terms of two modular squeezed quadratures:
\begin{align}
    c_1 c_2 [\op u, \op v]
        &=
            \left[
               \fracpart{\q_{c_1}}{ c_1 \perx }^{c_1 \perx/4} ,
               \fracpart{\p_{c^{-1}_2} }{ c_2 2\pi/\perx }^{c_2 \pi/a}
            \right]
        =
            0
                \,.
\end{align}
This expression identifies a family of commuting operators within a bosonic mode and means that pairs of modular squeezed quadratures can be measured with no back action between them.
The stretched Zak basis defined in \cref{eq:stretchedbasis1} is composed of the joint eigenstates for $c_1 = 1$ and $c_2 = a/b$. The commutator expression above provides a foundation for more general Zak bases, although we do not pursue them here.

\section{The GKP code} \label{sec:gkpgeneral}

\subsection{Zak description of the GKP code}
\label{sec:gkpstates}

The Gottesman-Kitaev-Preskill (GKP) code embeds a qubit into a rotor or a bosonic mode~\cite{gottesman_encoding_2001}, although more exotic Hilbert spaces can support it after appropriate generalization~\cite{albert_robust_2020}. (It is worth noting that the procedure involves a generalization of the Zak transform.) The code is constructed by considering a pair of momentum and position shift operators%
\footnote{
    One could consider more constructions by promoting $\X(a)$ and $\Z(b)$ to a
    pair of displacement operators instead~\cite{hanggli_enhanced}.
}
for some real values $a$ and $b$ such that
\begin{align}
\label{eq:conjugateops}
    \Z (a) \X (b)
        =
            \X (b) \Z (a)
    \,.
\end{align}
The parameter choice $ a b = 2 \pi K $ encodes a logical subspace of dimension $K$. We work with qubits, choosing $ a = 2 \alpha $ and $b = 2 \pi / \alpha $. The GKP codespace is defined by the stabilizer group
$
    \mc S
        =
            \langle
                \X ( 2\alpha ), \Z( 2 \pi / \alpha )
            \rangle
$.
One can then use the exponentiated canonical commutation relations to verify that odd powers of $\X(\alpha)$ and $\Z(\pi / \alpha)$ give pairs of anti-commuting operators that commute with $\mc S$ (\emph{i.e.},~they are logical). Even powers are in $\mc S$ instead. GKP states in the Hilbert space of the rotor or a bosonic mode are obtained by electing that the stabilizer generators are a pair of phase translation and momentum kicks. We get the GKP code in a bosonic mode when the pair of operators are interpreted as position and momentum shifts in a bosonic mode.

A stabilizer measurement for the GKP code is a simultaneous measurement of modular variables. This can be phrased nicely in the Zak framework using modular operators with natural periodicity $2\alpha$ (set $\perx=2\alpha$ in \cref{eq:zakperopU,eq:zakperopV,eq:zakperopUmod,eq:zakperopVmod}). Using \cref{eq:modularshifts} and the periodicity relation $T_U (2\alpha) = P_V (-2\alpha)$, the stabilizer group is expressed in terms of modular operators
\begin{align}
\label{eq:gkpstabgroup}
    \mc S
        =
            \langle
                \PVV(-2\alpha),
                    \PUU(2\pi/\alpha)
            \rangle
        =
            \langle
                e^{ - 2 i \alpha \op v },
                    e^{ i \frac{2\pi}{\alpha} \op u }
            \rangle
                \,,
\end{align}
which commute because $\op{u}$ and $\op{v}$ do. Expressing $\mc S$ this way makes it apparent that codespace will be spanned by simultaneous $\op u$ and $\op v$ eigenstates, or, in other words, by Zak vectors $\ket{u,v}$ with period $2\alpha$\blk. We can find the computational basis by looking for $+1$ eigenvalues of the stabilizer group
\begin{align}
\label{eq:codespace_1}
    \PVV(- 2\alpha)
        \ket{u,  v}
            &=
                e^{- 2 i \alpha v}
                    \ket{u,  v}
                        =
                            \ket{u,  v}
                                \,, \\*
\label{eq:codespace_2}
    \PUU(2 \pi/\alpha)
        \ket{u,  v}
            &=
                e^{ \frac{2 \pi i u}{\alpha} }
                    \ket{u,  v}
                        =
                            \ket{u,  v}
                                \,.
\end{align}
There are two solutions: the GKP codewords
\begin{align}
\label{eq:gkpcodeword}
    \ket{\ell_{\GKP}}
        =
            \ket{\alpha \ell, 0}
                \,, \quad
                    \ell \in \{0,1\}
                        \,.
\end{align}
The logical Paulis are a pure phase and a pure translation in modular position, respectively,
\begin{align}
\label{eq:loggkpuv}
    \Z_L
        &=
            \PUU \pqty{ \frac{\pi}{\alpha} }
                \,, \\
    \X_L
        &=
            \TUU (\alpha)
                \,,
\end{align}
as well as their odd powers. In terms of position eigenstates, the GKP codewords are
\begin{align}
\label{eq:GKPstates}
    \ket{\ell_\GKP} =
		\sum_{n=-\infty}^\infty \qket{ (2n + \ell) \alpha }                 \quad \text{for} \quad
                    \ell \in \{0,1\}
                        \,.
\end{align}
In position-space, the real number $\alpha$ sets the spacing of the grid where GKP states have support. For the square-lattice GKP code $\alpha = \sqrt{\pi}$; here, we leave it arbitrary.

We can apply the same reasoning to the more general choice of the stabilizer group for an asymmetric GKP code of dimension $K$,
$
    \langle
        \Z(a), \X( 2 \pi K/ a )
    \rangle
$.
It suffices to pick a set of modular operators with period $a$, so that we can use $ \Z(a) = \PVV(-a) $. To find the codespace, we start by looking for
$
    \ket{\bar u, \bar v}
$
such that
$
    \PVV(-a)
        \ket{\bar u, \bar v}
            =
                \ket{\bar u, \bar v}
$,
\emph{i.e.},~eigenstates of $\op v$ with value $ \bar v = 2 \pi n / a $, for $n\in \integers$. But the periodicity of the Zak basis in the second modular variable,
$
    \ket{\bar u, 2\pi / a}
        =
            \ket{\bar u, 0}
$,
ensures that the codewords are of the form $ \ket{\bar u, 0} $. Thus, it is always possible to pick a Zak basis (the one with period $a$) such that the codewords will be $0$-eigenstates of modular momentum. One can use the other eigenvalue equation and the quasi-periodicity in the first modular variable to show that the codespace is spanned by
$
    \ket{\bar \ell}
        =
            \ket{ \frac{a}{K} \ell, 0 }
$,
for
$\ell \in 0,\dots,K-1$. We have encoded a qudit of dimension $K$ by introducing $K$ bins in modular position with width $a/K$. We set $K=2$ from now on.

\begin{figure}[t]
    \includegraphics[width=0.35\textwidth]{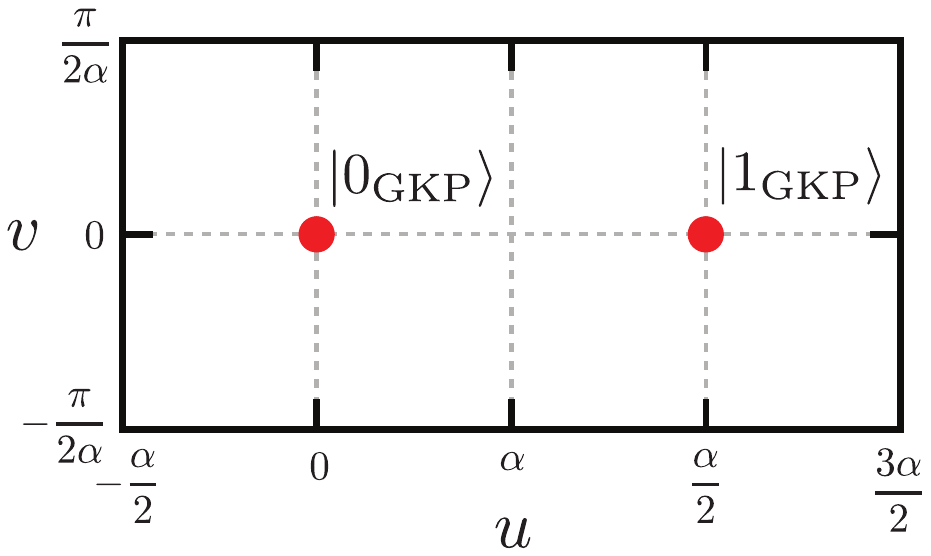}
    \caption{%
        Modular wavefunctions for GKP Pauli-$Z$ eigenstates $\ket{\ell_\GKP}$
        are $\delta(u - \alpha\ell,v)$. They are each represented as a single
        dot in a fundamental Zak domain $\zakdom$ with $a =
        2\alpha$~\cite{ketterer_quantum_2014}. Note that the dot color is
        consistent with the convention in \cref{fig:zakvacuuma}.
        \label{fig:centeringalpha}
    }
\end{figure}

\subsection{GKP error correction in the Zak picture}
\label{sec:ongrid}

GKP error correction (EC) projects a CV state into the GKP code subspace spanned by the GKP computational basis codewords. Error correction is not the only useful application of the error correction gadget: When the input state is known, we may also use GKPEC as an outcome-dependent projection into the codespace, so that we can post-select a state that is close enough to a target logical state such as a magic state~\cite{baragiola_allgaussian_2019}. In this section, we explore GKPEC and its connection to the Zak transform.

Let us consider the circuit for GKP syndrome extraction~\cite{gottesman_encoding_2001}:
\begin{equation}
\label{eq:GKPSMcircuit}
    \centerline{
    \Qcircuit @C=1em @R=1em{
        \lstick{\text{(out)}}
        & \gate{F} \qw
            & \ctrl{2}
                & \gate {F^\dag} \qw
                    & \ctrl{1}
                        & \qw
                            & \text{ (in)}
                                \\
        \lstick{\qbra{s}}
        & \qw
            & \qw
                & \qw
                    & \targ{-1}
                        & \qw
                            & \rstick{\hspace{-1em}\ket{+_\GKP}}
                                \\
        \lstick{\qbra{t}}
            & \qw
                & \targ{-2}
                    & \qw
                        & \qw
                            & \qw
                                & \rstick{\hspace{-1em}\ket{+_\GKP} \,.}
    }
}
\end{equation}
The circuit is read right-to-left,%
\footnote{
We find this notation more convenient than the conventional left-to-right because it preserves ordering when translating circuits into sequences of operators.}
and we represent a measurement in the position basis by sticking bras directly into the circuit's wires (\emph{e.g.},~$\bra{s}[q]$ is a position-quadrature measurement with outcome $s$). The boxed $F$ and $F^{\dag}$ are the Fourier transform and its adjoint respectively (note that $\F \ket{x}[q] = \ket{x}[p]$). We refer to the operator represented by this circuit as $ \op \Pi (s,t) $, where $s$ and $t$ are the outcomes of the two homodyne measurements. Although the syndromes can take on any real number, the circuit is only sensitive to shifts in modular position and modular momentum due to the periodicity in the GKP ancilla. This is by design, since measuring the integer parts would reveal logical information about the state.

The syndrome-extraction circuit can be rewritten in terms of a projector into GKP codespace,
$\gkpproj = \sum_{\ell} \ketbra{\alpha \ell, 0}$, that has been shifted according to the syndromes~\cite{baragiola_allgaussian_2019}
\begin{align}
    \op \Pi (s,t)
        &=
            \X\pqty{\util} \Z\pqty{\vtil}
                \gkpproj
                    \Z^{\dag}\pqty{\vtil} \X^{\dag}\pqty{\util}
                        \,,
\end{align}
where
\begin{align}
    \util
        \coloneqq
            \sfra \quad
                \text{and}
                    \quad
                        \vtil \coloneqq \tfra
                            \,.
\end{align}
As anticipated, the shifted projector~$\op \Pi (s,t)$ depends only on these fractional parts of the syndromes, which are confined to a patch $\gdom$ that defines the set of correctable shifts on an ideal GKP state,
\begin{equation}
\label{eq:PG}
    \gdom
        =
            [-\alpha/2,\alpha/2)
                \times
                    [-\pi/2\alpha,\pi/2\alpha)
                        \,.
\end{equation}
This patch, with area $\pi$, is only \emph{half} the size of the fundamental Zak domain $\mathcal{P}$, \cref{eq:zakdom}. In the Zak picture, the GKP syndrome-measurement operator can be written in terms of modular phases and translations. Using \cref{eq:modularshifts}, it is
\begin{align}
    \synst
        &=
            \label{eq:modsynd}
                \TUU(\util)
                    \TVV(\vtil)
                        \gkpproj
                            \TVV(-\vtil)
                                \TUU(-\util)
                                    \\
        &=
            \label{eq:modsyndkb}
                \sum_{\ell=0,1}
                    \ket{\util + \alpha \ell , \vtil}
                        \bra{\util + \alpha \ell , \vtil}
                            \,,
\end{align}
where $\ell$ labels the GKP logical state as in \cref{eq:GKPstates}, and the modular phases have cancelled.

After syndrome extraction, GKP error correction returns the conditional state to the codespace using an outcome-dependent recovery operation composed of  displacements in position and momentum by $-\util$ and $-\vtil$,
\begin{align}
    \op{\mathcal{R}}(\util,\vtil)
    &\coloneqq
        \Z(-\vtil)
            \X(-\util) \\
                &=
                    \PUU(-\vtil)
                        \TVV(-\vtil)
                            \TUU(-\util)
                                \,.
\end{align}
Composing syndrome extraction and recovery gives the Kraus operator for GKP error correction, presented here in Zak form
\begin{align}
\label{eq:GKPEDdecomp}
    \gkpec
        &=
            \op{\mathcal{R}}(\util,\vtil) \synst
                \\
        &=
            \PUU(-\vtil)
                \gkpproj
                    \TVV(-\vtil)
                        \TUU(-\util)
                            \,.
\end{align}
The modular phase operator $ \PUU(-\vtil) $ generates a logical rotation around the $Z$-axis when acting on a GKP state,
\begin{align}
\label{eq:phasetorotation}
    \PUU(-\vtil)
        \ket{\ell_{\GKP}}
            &=
                e^{-i \alpha \ell \vtil }
                    \ket{\ell_{\GKP}}
                        \,, \nonumber \\
            &=
                e^{i \phi}
                    \op{R}_L(2 \vtil \alpha)
                        \ket{\ell_{\GKP}}
                            \,,
\end{align}
where $\op{R}_L(\theta) \coloneqq e^{i \frac{\theta}{2} \op{\sigma}_z}$ is a qubit rotation operator on the GKP subspace, and $\phi$ is a trivial phase.

Performing GKP error correction on an arbitrary, pure CV state $\ket{\psi}$ gives a conditional state in the GKP codespace
\begin{align}
\gkpec
    \ket{\psi}
        &=
            \bar{c}_0 \ket{0_\GKP} + \bar{c}_1 \ket{1_\GKP}
                \,,
\end{align}
with unnormalized complex amplitudes given by wavefunctions in the Zak representation
\begin{align}
\label{eq:ECcoeffs}
    \bar{c}_\ell
        =
            e^{-i \alpha \ell \vtil }
                \psi\pqty{\util+\alpha\ell,\vtil}
                    \,.
\end{align}
Each amplitude is a Zak transform of the position-shifted initial state,
$
    \psi(\util+\alpha \ell,\vtil)
        =
            \bra{\util + \alpha \ell,\vtil}
                \ket{\psi}
        =
            \bra{\util ,\vtil}
                \op{X} ^\dag ( \alpha \ell )
                    \ket{\psi}
$,
with an additional phasing in $\bar c_1$ due to the final logical rotation, \cref{eq:phasetorotation}.

\subsection{Mapping a CV state to a qubit state}
\label{sec:CVtoqubit}

A qubit state $\op{\rho} = \sum_{\ell,\ell'} \rho_{\ell \ell'} \ketbra{\ell}{\ell'}$ is encoded into a CV state in the GKP code
\begin{align}
    \rhop_\GKP
        =
            \sum_{\ell,\ell'}
                \rho_{\ell \ell'}
                    \ket{\ell_\GKP}
                        \bra{\ell'_\GKP}
                            \,.
\label{eq:sigmagkp}
\end{align}
The inverse question of how to construct a map from any CV state into the Bloch sphere in a way that is compatible with the GKP code is much subtler, because the map above is not uniquely invertible, and we have yet to define the notion of compatibility with the GKP code. We construct this inverse map is by averaging over states that are assumed to well approximate GKP codewords over the set of all correctable errors. The average will decrease the dimensionality of the state, and at the same time, approximate GKP states will be associated to qubit states that have good overlap with the intended qubit state with matrix elements $\rho_{\ell \ell'}$.

Consider a CV state $\rhop$, assumed to approximate $\rhop_\GKP$ well. An ideal GKP state is error correctable whenever it undergoes a combination of small quadrature displacements $\X(\tilde{u})\Z(\tilde{v})$ with $\tilde{u}, \tilde{v}$ confined to the patch $\gdom$, \cref{eq:PG}. We expect the same to hold, approximately, for $\rhop$. Thus, the states $\rhop$ and
$
\rhop(\tilde{u},\tilde{v})
    =
        \Z^{\dag}(\tilde{v}) \X^{\dag}(\tilde{u})
            \rhop
                \X(\tilde{u}) \Z(\tilde{v})
$
are logically equivalent, up to how well $\rhop$ approximates $\rhop_\GKP$. We integrate the small shifts away and note that, for reasonably good approximate GKP states, we can define qubit matrix elements as if \cref{eq:sigmagkp} were to hold and expect them to be reasonably close to the target matrix elements $\rho_{\ell \ell'}$. Specifically,
\begin{align}
    \tilde\rho_{\ell \ell'}
        &\coloneqq
            \int_{ \gdom }
                d \tilde{u}\, d \tilde{v}  \,
                       \bra{\ell_{\GKP}} \Z^{\dag}( \tilde{v} ) \X^{\dag}( \tilde{u} )
                            \rhop
                                \X( \tilde{u}) \Z( \tilde{v} )
                                    \ket{\ell'_{\GKP}}
                                        \,.
\label{eq:logstate}
\end{align}
The emergence of this half domain is the fundamental reason why, later on, we will find it convenient to use a stretched Zak basis with area $\pi$. Using $\ket{\ell_{\GKP}} = \ket{\alpha \ell,0}$ and acting with translation operators on the Zak kets, we can also express \cref{eq:logstate} as
\begin{align}
\label{eq:commonsensequbituvrho}
    \tilde \rho_{\ell \ell'}
        &=
            \int_{ \gdom  }
                d \tilde{u}\, d \tilde{v} \,
                    \bra{\tilde{u} + \alpha \ell , \tilde{v}}
                        \rhop
                            \ket{\tilde{u}+\alpha \ell', \tilde{v}}\\
         &=
            \sum_{n}\text{Pr}_n \int_{ \gdom  }
                d \tilde{u}\, d \tilde{v} \,
                \psi_n\pqty{
                        \tilde{u}
                        + \alpha \ell,\tilde{v}
                    }^{*}
                        \psi_n\pqty{
                            \tilde{u}
                            + \alpha \ell',\tilde{v}
                        }
    \,,
\end{align}
which we have cast in terms of Zak transforms in the second line by expressing $\rhop$ in diagonal form,
$
\rhop
    =
        \sum_n \text{Pr}_n \ket{\psi_n}\bra{\psi_n}
$,
where $\text{Pr}_n$ is the mixture probability of state $\ket{\psi_n}$.

Although designed for approximate GKP states, the above procedure can be used to find a ``logical'' GKP state associated to any CV state,
\begin{align}
\label{eq:logqubit}
    \op \rho_L[\rhop]
        \coloneqq
            \sum_{\ell, \ell'} \tilde \rho_{\ell \ell'} \ketbra{\ell}{\ell'}
                \,,
\end{align}
with qubit matrix elements given by \cref{eq:logstate}. When $\rhop$ is an approximate GKP state, the fidelity between $\op \rho_L$ and the intended qubit state increases monotonically (\emph{i.e.},~$\tilde \rho_{\ell \ell'}$~approaches $\rho_{\ell \ell'}$) with the quality of the approximation, and so does the purity. In fact, in this picture, most approximate GKP states encode mixed qubits. Interestingly, the qubit state defined by \cref{eq:logstate} coincides with that obtained by a partial trace over the gauge subsystem in the ``partitioned-position'' (PP) subsystem decomposition~\cite{pantaleoni_modular_2020,pantaleoni_hidden_2021}---as we will show in \cref{sec:uvdec}. A thorough investigation of PP subsystem qubits encoded by approximate GKP states and other common resource states in continuous-variable quantum computing can be found in Ref.~\cite{pantaleoni_subsystem_2021}.

\section{Modular variable subsystem decomposition} \label{sec:modvarSSD}

In \cref{sec:CVtoqubit}, we have given procedures to associate to any CV state a qubit state whose matrix elements involve partial integrals of the state's Zak transform over the $\pi$-area patch $\gdom$ in \cref{eq:PG}. In this section, we revisit this idea from the perspective of a bosonic subsystem decomposition (SSD), introduced in Ref.~\cite{pantaleoni_modular_2020}. This procedure, which is not uniquely defined, decomposes the infinite-dimensional Hilbert space of the CV mode into that of  a virtual qubit and  another full CV mode, $\HCV \simeq \complex^2 \otimes \HCV $, by means of a change of basis.  With this decomposition, the qubit subsystem carries the logical information, and the CV subsystem, referred to as the gauge mode, carries no logical information. Below, we construct a \emph{modular variable subsystem decomposition} (MV SSD) in the form $\text{(qubit)} \otimes \text{(mode)}$ using the Zak-transform tools introduced above.  Ultimately, we find that the tensor-product structure of this SSD is the same as that for the partitioned-position SSD~\cite{pantaleoni_modular_2020, pantaleoni_subsystem_2021}, except that the gauge mode is expressed in the stretched Zak basis. However, the MV SSD is specifically suited to describe GKP error correction, because the GKP error recovery procedure extracts syndrome information by measuring modular position and momentum.

\subsection{Construction}
\label{sec:uvdec}

The Zak basis is well suited for a subsystem-decomposed description of periodic states.
In analogy with the procedure we followed when defining the partitioned-position SSD~\cite{pantaleoni_modular_2020},  we construct the SSD by decomposing an operator diagonal in a basis for $\HCV$: the modular position operator in the Zak basis (as opposed to the position operator in the position basis, which is the strategy we used for the partitioned-position decomposition). Using Aharonov's modular variables, \cref{eq:Aharonov}, we have
\begin{align}
\label{eq:aha2alpha}
    \q
        &=
            \op{u} + 2 \alpha \op{m}
                \,, \\
        \p
            &=
                \op{v} + \frac{\pi} {\alpha} \op{n}
                    \,,
\end{align}
where we choose a bin size of $2\alpha$.%
\footnote{%
    In the partitioned-position
    SSD~\cite{pantaleoni_modular_2020,pantaleoni_subsystem_2021} the bin size
    is $\alpha$; thus the operator $\op{u}$ here is different from $\op{u}$
    there.
}
The domains of
$
\op{u}
    =
        \fracpart{\q}{2\alpha}^{\alpha/2}$,
$
\op{v}
    =
        \fracpart{\p}{\pi/\alpha}^{\pi/2\alpha}
$,
$
\op{m}
    =
        \closestint{\q}{2\alpha}^{\alpha/2}
$ and
$
\op{n}
    =
        \closestint{\p}{\pi/\alpha}^{\pi/2\alpha}
$ are chosen such that the modular variables lie in a fundamental, $2\pi$-area Zak patch
\begin{align}
    \mathcal{P}
        =
            \Big[
                - \frac{\alpha}{2},\frac{3\alpha}{2}
            \Big )
                \times
                    \Big[
                        -\frac{\pi}{2 \alpha},\frac{\pi}{2\alpha}
                    \Big)
                        \,.
\end{align}
A useful choice is $\alpha = \rpi$, which makes the rectangle exactly the union of two identical squares with sides $\rpi$---in fact, the figures in this article reflect this choice.

We define a set of \emph{logical} and \emph{gauge} operators as simple functions of the modular operators, where the parameter $\alpha$ is taken as the half-period of the GKP codewords in~\cref{eq:gkpcodeword} (this where the construction becomes compatible with the GKP code):
\begin{align}
\label{eq:ugauge}
    \uGop
        &\coloneqq
            \fracpart{\op{u}}{\alpha}
                \,, \\
\label{eq:vgauge}
    \vGop
            &\coloneqq
                \op{v}
                    \,, \\
\label{eq:logop}
    \l
        &\coloneqq
            \frac{\closestint{\op{u}}{\alpha}}{\alpha}
                \,,
\end{align}
omitting the centering, which is understood throughout the rest of the article as $\alpha/2$ for the modular position truncation, and $\pi/2\alpha$ for modular momentum one. The \emph{which-patch} operator $\l$ indicates whether a modular position eigenstate in the original patch $\zakdom$ is closer to $0$ or to $\alpha$.
\begin{figure}[t]
    \includegraphics[width=0.4\textwidth]{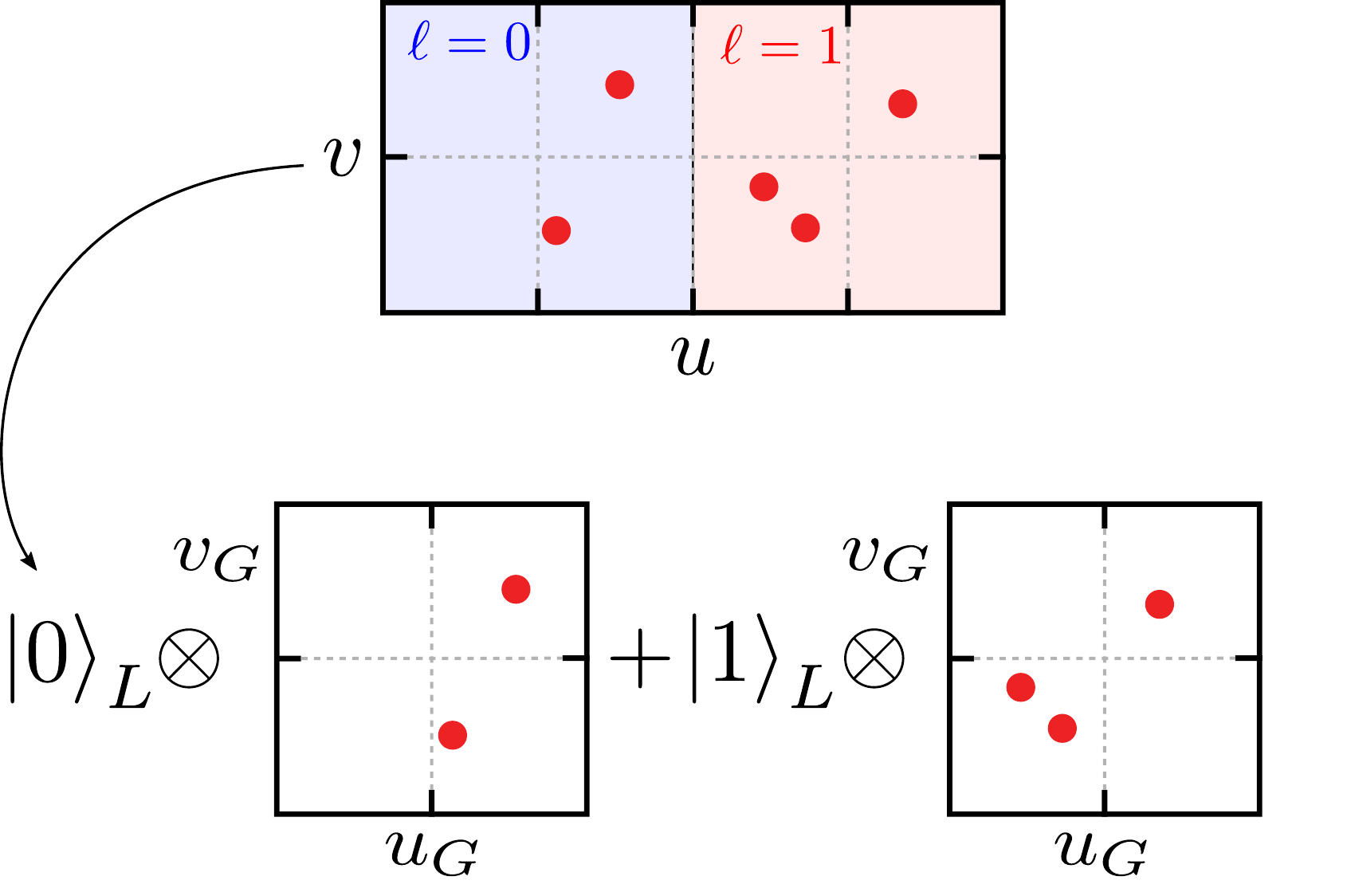}
    \caption{%
        Graphical depiction of the MV SSD. Top:~A linear combination of
        modular-shifted GKP states,
        $
        \sum_i \TUU({\util}_i)\TVV({\vtil}_i)\ket{0_\GKP}
            =
                \sum_i\ket{\util_i,\vtil_i}
        $,
        represented in the fundamental Zak patch
        $\zakdom$. Bottom:~After the change of basis into the MV SSD, the
        smaller gauge patch $\gdom$ no longer carries left/right information,
        which is instead transferred into the logical index spanning a basis
        for $\complex^2$. The final state is then decomposed as a linear
        combination of
        $
            \ket{j}[L]
                \otimes
                    \ket*{
                        \fracpart{\util_i}{\alpha}^{\alpha/2},\vtil_i
                    }[G]
        $.
    }
\label{fig:patch}
\end{figure}

The inverse relation
\begin{align}
\label{eq:u_op_decomposed}
    \op{u}
        &=
            \uGop + \alpha \l
                \,
\end{align}
reveals that modular position can be fully reconstructed from $\l$ and $\uGop$. Thus, since the operators $\l$, $\uGop$, and $\vGop$ commute, their eigenvalues label a basis for the CV mode, since the eigenvalues of $\op{u}$ and $\op{v}$ do. The change of basis induced by the operator decomposition above is
\begin{align}
\label{eq:uvchangeofbasis}
    \ket{u,v}
        &=
            \ket{
                \closestint{u}{\alpha}
                    + \fracpart{u}{\alpha},v } \\
        &=
            \ket{ \closestint{u}{\alpha} / \alpha }[L]
                \otimes
                    \ket{ \fracpart{u}{\alpha},v }[G]
                        \,,
\end{align}
where the second line introduces a tensor-product structure between a virtual logical qubit and a virtual gauge subsystem. This tensor-product structure defines a subsystem decomposition. The eigenvalues of the gauge subsystem lie in the Voronoi cell of the symplectically dual lattice as defined in Gottesman \emph{et al.}~\cite{gottesman_encoding_2001}, which is exactly the $\pi$-area patch $\mathcal{P}_G$ defining the set of correctable shifts on an ideal GKP state, \cref{eq:PG}.

A Zak-basis state in the left square of the full patch $\mathcal{P}$ is associated with the state $\ket{0}[L]$ of the logical subsystem and a right square with $\ket{1}[L]$, as illustrated in~\cref{fig:patch}. The gauge subsystem lies in a stretched Zak space $\gdom$ with area $\pi$, and is the Hilbert space of a (virtual) bosonic mode. Thus, we refer to the gauge subsystem as a \emph{gauge mode}~\cite{pantaleoni_modular_2020}. The periodicities of the gauge eigenstates are given by setting $\perx = \alpha$ and $\pery = 2 \alpha$ in \cref{eq:abperiods}:
\begin{align}
\label{eq:gaugeperU}
    \ket{ u + \alpha,v}[G]
        &=
            e^{ - i 2 \alpha v } \ket{u,v}[G]
                \,, \\
\label{eq:gaugeperV}
    \ket{ u,v + \pi/\alpha }[G]
        &=
            \ket{u,v}[G]
                \,.
\end{align}
We use the $\alpha$-quasi-periodicity above to give an alternate form for the change of basis by plugging $\fracpart{u}{\alpha} = u - \closestint{u}{\alpha}$ into \cref{eq:uvchangeofbasis},
\begin{align}
\label{eq:uvchangeofbasisnew}
    \ket{u,v}
        =
            e^{2 i v \closestint{u}{\alpha} }
                \ket{ \closestint{u}{\alpha}/\alpha }[L]
                    \otimes
                        \ket{ u , v }[G]
                            \,.
\end{align}
Because $u\in[-\alpha/2,3\alpha/2)$, the gauge Zak state above is not centered, and the phase multiplying the state may or may not cancel depending on which patch $u$ lies in.

As far as the logical-qubit subsystem is concerned, since $\l \ket{\ell}[L] = \ell \ket{\ell}[L]$, the operator $\l$ is the projector into $\ket{1}[L]$ generating a logical Pauli-$Z$ by exponentiation as $\Z_L = e^{i \pi \l}$.

The completeness relation of the full mode with respect to the SSD gives the expressions for the gauge and logical identities. We can write the completeness for the Zak basis as a sum of integrals over the left and right patches,
\begin{align}
    \op \id_{\text{CV}}
        =
            \sum_{\ell \in 0,1}
                \int_{
                    -\frac{\alpha}{2} + \alpha \ell
                }^{
                    \frac{\alpha}{2} + \alpha \ell
                } du
                    \int_{ -\pi/2\alpha }^{ \pi/2\alpha }
                        \! \! \! dv\,
                            \ket{ u+\alpha \ell,v }
                                \bra{ u + \alpha \ell,v }
                                    \,,
\end{align}
and use the change of basis in \cref{eq:uvchangeofbasis}, as well as the fact that $\fracpart{u}{\alpha} = u$ holds in the integration domain. Thus,
\begin{align}
\label{eq:uvcompleteness}
    \op\id_{\text{CV}}
        =
            \sum_{\ell\in{0,1}}
                \ket{\ell}[L]\bra{\ell}[L] \otimes
                        \intgdom du\, dv\,
                            \ket{u,v}[G]\bra{u,v}[G]
        =
            \id_{L} \otimes \id_{G}
                \,.
\end{align}
This identifies the completeness relations for the logical subsystem and the gauge mode. Using \cref{eq:uvcompleteness}, we can write any CV state in the Zak subsystem basis as
\begin{align}
\label{eq:anyrho}
    &\rhop
        =
            \sum_{\ell \in{0,1}}
                \sum_{\ell'\in{0,1}}
                    \int_{\gdom} \! \! du\, dv\,
                        \int_{\gdom} \! \! du'\, dv'\,
                            \nonumber \\
            &\quad \times
                \rho_{\ell \ell'}(u,u',v,v')
                    \ket{\ell}[L] \bra{\ell'}[L]
                        \otimes
                            \ket{u,v}[G] \bra{u',v'}[G]
\end{align}
with matrix elements
\begin{align}
\label{eq:rhoZakelements}
    \rho_{\ell \ell'}(u,u',v,v')
        \coloneqq
            (\bra{\ell}[L] \otimes \bra{u,v}[G])
                \rhop
                    (\ket{\ell'}[L] \otimes \ket{u',v'}[G])
                        \,.
\end{align}
Since the SSD defines a complete basis over the full mode, it can also be used to write other CV objects, such as operators or channels, in the logical-gauge tensor-product basis. Once a suitable method for extracting a logical-qubit state is defined, such decompositions can be useful techniques for identifying the logical action of such objects.

The tensor-product decomposition allows us to revisit the question in \cref{sec:CVtoqubit}: what is the logical qubit associated with a CV state? A straightforward way to extract a logical-qubit density matrix from a CV state $\rhop$ is by taking a trace with respect to the gauge subsystem,
\begin{align}
\label{eq:uvgaugetrace}
    \rhop_L[\rhop]
        =
            \Tr_G \rhop
        =
            \intgdom du\, dv\,
                \bra{u,v}[G] \rhop \ket{u,v}[G]
                    \,.
\end{align}
The gauge mode is defined over a patch $\gdom$, so taking this subsystem trace is equivalent to the map in \cref{eq:logstate}, and the qubit state is the identical to the one in \cref{eq:logqubit}, which is why we use the same notation, $\rhop_L[\rhop]$. This equivalence is straightforward: Starting from \cref{eq:commonsensequbituvrho} for the qubit matrix elements,
\begin{align}
\label{eq:gaugetracederivation}
    \tilde \rho _{\ell \ell'}
        &=
            \intgdom
                du\, dv\,
                    \bra{u + \alpha \ell , v}
                        \rhop
                            \ket{u+\alpha \ell',v}
        \\ &=
            \intgdom
                du\, dv\,
                    \big(\bra{\ell}[L] \otimes \bra{u, v}[G] \big)
                        \rhop
                           \big( \ket{\ell'}[L] \otimes \ket{u,v}[G] \big)
        \\ &=
            \intgdom
                du\, dv\,
                \rho_{\ell \ell'}(u,u,v,v)
        \\ &=
            \bra{\ell}[L]
                (\Tr_G \rhop)
                    \ket{\ell'}[L]
                        =
                            (\Tr_G \rhop) _{\ell \ell'}
    \,,
\end{align}
as anticipated. The third form uses the matrix elements of~$\rhop$, \cref{eq:rhoZakelements}, showing that this partial trace can be interpreted as integration of the ``gauge diagonal'' of these matrix elements (\emph{i.e.},~$u=u'$, $v=v'$) over the gauge patch, for each of the four choices of logical elements~$\ell,\ell'$.

Ideal GKP states are not the only states whose gauge trace is a pure state: Any product state of the form $\op \rho = \ketbra{\psi}{\psi}[L][L] \otimes \op{\rho}_G$ encodes a pure logical state, even  when $\op{\rho}_G$ is mixed, allowing for mixed-state quantum computation~\cite{Lau_mixed2017,Lau_mixed2019}.

It should be noted that, while we usually interpret the gauge mode as a subsystem containing redundant infor-mation, we are not in the same situation that is set up for subsystem codes. Overall, error-correctable shifts that we have referred to as ``gauge operators'' must be detected via a syndrome measurement and corrected. After all, the GKP code is a subspace code, and remains as such, despite our subsystem construction rephrases the code in terms of two subsystems. The interpretation of subsystem codes in terms of subsystems is not a new idea: In fact, it was introduced in
Refs.~\cite{knill1997theory,knill2000theory}---see, specifically, Theorem 3.5 in the former reference, where the ``syndrome subsystem'' plays the same role as what we have called ``gauge subsystem''.

\subsubsection{Connection to the partitioned-position SSD}
\label{sec:PPSSDconnection}

We can draw an explicit connection (in fact, an equivalence) between the MV SSD presented here, \cref{eq:ugauge,eq:vgauge,eq:logop}, and the partitioned-position SSD in Refs.~\cite{pantaleoni_modular_2020,pantaleoni_subsystem_2021}. We do so by first comparing their logical operators. Consider the partitioned-position logical operator, $\l_\text{PP}$, defined as
\begin{align}
    \l_\text{PP}
        =
            \fracpart{
                \frac{
                    \closestint{ \q }{ \alpha }
                }{\alpha}
            }{2}
                \,,
\end{align}
where $\fracpart{n}{2}$ gives the parity of the integer $n$. In terms of Aharonov's modular variables, \cref{eq:aha2alpha}, we have
\begin{align}
   \l_\text{PP}
        =
            \fracpart{
                \frac{
                    \closestint{
                        \op u + 2 \alpha \op m
                    }{ \alpha }
                }{ \alpha }
            }{2}
        =
            \fracpart{
                \frac{
                    \closestint{
                        \op u
                    }{ \alpha }
                }{ \alpha }
                    + 2 \op m
            }{2}
        =
            \frac{
                \closestint{
                    \op u
                }{ \alpha }
            }{ \alpha }
                \,,
\end{align}
which is exactly~\cref{eq:logop}. Since $\l_\text{PP}$ and the $\l$ in \cref{eq:logop} are the same operator, they define the same logical-qubit subsystem. As a result, the two SSDs are related by a change of basis on the gauge modes. The change of basis is a Fourier series of the partitioned-position gauge mode states (an inverse Zak transform), $\ket{m,u}[G]$,
\begin{align}
\label{eq:pptozak}
    \ket{\ell}[L] \otimes \ket{u,v}[G]
        =
            \ket{ \ell }[L] \otimes
                \sqrt{\frac{\alpha}{\pi}}
                    \sum_{m \in \mathbb{Z}}
                        e^{ i 2 \alpha m v}
                            \ket{ m, u }[G,\text{PP}]
                                \,,
\end{align}
where $u \in [-\alpha/2,\alpha/2)$. The important feature to highlight is that the tensor-product structure of the SSDs is the same.

The reader interested in the analysis of approximate resource states for protocols that rely on finite-energy GKP states, as well as cluster states, can leverage this equivalence. For example, the logical analysis in Refs.~\cite{pantaleoni_subsystem_2021,pantaleoni_hidden_2021} use the partitioned-positioned decomposition. In this picture, states have the same logical subsystem of the modular-variable decomposition presented here. Further, one can obtain the gauge wavefunctions in the modular-variable decomposition via a Zak transform of the gauge mode alone, as shown in \cref{eq:pptozak}.

\subsection{Phases and translations in the subsystem-decomposed picture}
\label{sec:uvshifts}

States and operators can be decomposed and interpreted in the modular variable subsystem description. We focus on the quadrature-shift operators $\X(t)$ and $\Z(t)$, since they form an operator basis~\cite{serafini_quantum_2017}, meaning that their subsystem decompositions can be used to decompose unitaries and more general quantum operations on the CV mode. Moreover, specific shifts implement logical Paulis on GKP states, and small shifts describe correctable errors.

The periodicities of the modular gauge phase and translation operators are read from \cref{eq:stretchedopperiods}, choosing $\perx = \alpha$ and $\pery = \pi/\alpha$,
\begin{align}
\label{eq:gaugeopperiods}
    \TUG( \alpha ) &= \PVG( -2\alpha )
        \\
            \TVG( \pi/\alpha ) &= \op \id_G
                \,.
\end{align}
The winding number of $u$-translations differs by a factor two with respect to the full-mode case \cref{eq:zakperopU}.

It is sufficient to consider how modular-phase and modular-translation operators in the original Zak patch $\zakdom$ decompose in the MV SSD, since we can use \cref{eq:xmodular,eq:zmodular} to reconstruct the quadrature shift operators. The subsystem decomposition leaves modular-momentum unchanged, so we immediately find that $\PVV(t)=\op \id_L \otimes \PVG(t)$ and $\TVV(t)=\op \id_L \otimes \TVG(t)$. The phase operator associated to modular position decomposes easily: $\PUU(t) = e^{ i ( \alpha \l + \uGop ) t } = e^{ i\alpha \l t } \otimes \PUG(t)  $, where we used the decomposition of the modular position operator, \cref{eq:u_op_decomposed}, and $\PUG(t) = e^{i \uGop t }$. We find the decomposition of momentum shift operator $\Z(t)$, from \cref{eq:zuv},
\begin{align}
\label{eq:z_dec}
    \Z(t)
        =
            e^{ i\alpha \l t } \otimes \TVG(t) \PUG(t)
                \,,
\end{align}
a tensor-product operator across the logical and gauge subsystems.

The MV SSD of $\TUU(t)$ is more complicated. We first split the translation into an integer and a fractional component,
$
    \TUU(t)
        =
            \TUU(\tinta) \TUU(\tfraa)
$.
The first term (on the left) is a logical bit flip depending on the bin-parity of the $t$ parameter:
$
    \TUU(\tinta)
        =
            \X_{L}^{\frac\tinta\alpha} \otimes \op \id_G
$.
The second term can be evaluated in the Zak basis using \cref{eq:uvchangeofbasisnew}. We obtain
\begin{align}
\label{eq:x_dec_derivation}
    &\TUU(\tfraa) \ket{ \alpha \ell + u, v }
        \nonumber \\
            &\quad=
                \X_L^{ \frac{ \closestint{u + \tfraa}{\alpha} }{\alpha} }
                    \TUG(\tfraa)
                        e^{ i \vGop \closestint{ \tfraa + \uGop }{ \alpha } }
                            \ket{\ell}[L] \otimes
                                \ket{ u , v }[G]
                                    \,.
\end{align}
Using the fact that the decomposed basis is complete, adding the remaining logical operation $\X_{L}^{\frac{\tinta}{\alpha}}$, and recalling that $\op{X}(t) = \TUU(t) $, the position-shift operator decomposes as
\begin{align}
\label{eq:x_dec}
    \op{X}(t)
        &=
            \X_{L}^{\frac{\tinta}{\alpha}}
                \X_L^{ \frac{ \closestint{ \op u_G + \tfraa }{\alpha} }{\alpha} }
                    \TUG(\tfraa)
                        e^{ i \vGop \closestint{ \tfraa + \uGop }{ \alpha } }
                            \,.
\end{align}
This is not a tensor-product operator. The factor
$
    \TUG(\tfraa)
        e^{ i \vGop \closestint{ \tfraa + \uGop }{ \alpha } }
$
acts only the gauge mode, phasing and shifting it in a nontrivial way.
However, there is a logical-gauge entangling factor,
$
    \X_L^{ \frac{ \closestint{ \uGop + \tfraa }{\alpha} }{\alpha} }
$,
also discussed in Ref.~\cite{pantaleoni_subsystem_2021}, whose purpose is to account for the logical bit flip that occurs when the value of $\tfraa$ is sufficiently large to translate a gauge modular position eigenstate outside of the half-patch, \emph{i.e.},~when $\alpha/2 \leq \tfraa + u < 3\alpha/2$.

Quadrature-shifted GKP states,
$
    \Z(\util) \X(\vtil)
        \ket{\ell_{\text{GKP}}}
$,
can be evaluated in their decomposed form with the expressions we derived in this section. They simplify greatly due to the fact that GKP state is a $0$-eigenstate of both $\uGop$ and $\vGop$.
In addition, when the displacements are small
we have
\begin{align}
\label{eq:smallshiftGKP}
    \X(\util)
        \Z(\vtil)
            \ket{\ell}[L] \otimes \ket{0,0}[G]
    &=
        e^{i \alpha \l \otimes \op v_G  }
            \ket{\ell}[L] \otimes \ket{\util, \vtil}[G]
                \,, \nonumber \\
        &\quad \quad
            \util,\vtil \in \gdom
                \,.
\end{align}
Small position shifts do not disturb logical information, while a momentum shift by $\vtil$ induces a logical rotation around the $Z$-axis of the Bloch sphere, $ e^{i \alpha \vtil \l} $. On the other hand, a well-known property of the GKP code is full protection against small shifts. Thus, we expect the GKP error correction gadget to apply a logical rotation in the opposite direction, so as to cancel this unwanted logical distortion. This is indeed the case, as we will see in the next section.

\subsection{GKP error correction in the subsystem-decomposed picture}\label{sec:uvprojdec}

The MV SSD is a change of basis in the Hilbert space of a bosonic mode that is compatible with the GKP code in the sense that ideal states are separable:
\begin{align}
\label{eq:SSDidealGKP}
    \ket{\psi_\GKP}
        =
            \ket{\psi}[L] \otimes \ket{0,0}[G]
                \,.
\end{align}
The gauge-mode state is $ \ket{0,0}[G] = \sum_m \ket{\alpha m + u}[q] = \ket{+_{\text{GKP}}}[G]
$. GKP states play a special role in this decomposition, by design, but we still have not fully discussed how their resilience with respect to small shifts fits into our paradigm. This discussion is the aim of this section.

\Cref{eq:smallshiftGKP} shows that a small position shift $\X(t)$, $t \in [-\alpha/2, \alpha/2)$ acts as an identity on the logical subsystem while inducing a small shift on the gauge mode. Thus, protection against a small position shift is well described in the decomposed basis: a random small shift induces no logical errors, manifests itself as syndrome information encoded by the gauge mode.

On the other hand, \cref{eq:smallshiftGKP} also shows that momentum shifts, $\Z(t)$, rotate the qubit subsystem around the $z$-axis of the Bloch sphere by an angle $\alpha t/2$, regardless of the magnitude $t$ of the shift. At first glance, this is a discrepancy with the error model of the GKP code, as one would expect small momentum shifts to act as logical identities just like for position shifts. From an error correction perspective, they are correctable in the same sense that small position shifts are. Because of this asymmetry in how small shifts are described in the SSD, we conclude that simply inspecting the logical subsystem may not be the most accurate figure of merit for the logical content of a codeword with respect to the GKP code.

This additional logical operation under \(\Z(t)\), however, is merely a logical artifact due to the choice of the decomposed basis. To remedy this, we propose modifying the expression for the gauge trace so that it will give a logical qubit that agrees with the procedure for GKP error correction and recovery, as discussed next.

First, note that since
$
    \ket{\ell_{\GKP}}
        =
            \ket{\ell}[L]
                \otimes
                    \ket{0,0}[G]
                    $,
the projector into the GKP codespace decomposes into a tensor product operator
$
\gkpproj
    =
        \sum_{\ell} \ketbra{\alpha \ell, 0 }
            =
                \id_{L} \otimes \ketbra{0,0}[GG]
$.
Given two measurement outcomes associated to the error correction procedure, $s$ and $t$, we write $\util = \sfra$ and $\vtil = \tfra$ and decompose GKPEC [\cref{eq:GKPEDdecomp}] as
\begin{align}
\label{eq:GKPECSSD}
    \gkpec
        &=
            (
                \id_{L}
                    \otimes
                        \ketbra{0,0}[GG]
            )
                \Z^{\dag}(\vtil)
                    \X^{\dag}(\util)
                        \\
        &=
            e^{-i \alpha \l \otimes \vtil_G}
                \ketbra{0,0}{ \util,\vtil }[GG]
                        \,.
\end{align}
In the MV SSD, GKPEC consists of a gauge-mode measurement, followed by a projection of the gauge subsystem into $\ket{0,0}[G]$---the gauge state of a perfectly encoded GKP qubit---, followed by an entangling operation.
The entangling operation
$
e^{-i \alpha \l \otimes \vtil_G}
$
is a counter-rotation that undoes the logical artifact appearing in small-shifted GKP states, in accordance with the fact that small shifts are correctable and bear no consequences on the logical information of a GKP state after error correction.

Performing GKP error correction on pure CV state $\ket{\psi}$ gives a conditional tensor-product state,
\begin{align}
\gkpec \ket{\psi}
    &=
        \big(
            \bar{c}_0 \ket{0}[L] + \bar{c}_1 \ket{1}[L]
        \big)
            \otimes
                \ket{0,0}[G]
                    \,,
\end{align}
which is an ideal GKP state according to \cref{eq:SSDidealGKP}. By expressing the state $\ket{\psi}$ in the Zak basis with expansion coefficients $\psi_\ell(u,v)$, the complex amplitudes above can be written as
$
    \bar{c}_\ell
        =
            e^{-i \alpha \ell \vtil} \psi_\ell(\util,\vtil)
$,
which is an alternate form for those in \cref{eq:ECcoeffs}.

More generally, we can construct the full GKP error correction channel from the Kraus operators~$\gkpec$:
\begin{align}
    \ecchan
        &\coloneqq
            \intgdom d\util\, d\vtil\,
                \eckraus \pqty{\util, \vtil}
                    \odot
                        \eckraus^\dag \pqty{\util, \vtil}
                            \\
    &=
        \intgdom d\util\, d\vtil\,
        (
            e^{-i \alpha \l \vtil  }
                \otimes
                    \ketbra {0,0} {\util,\vtil}[GG]
        )
            \nonumber \\
    &\qquad \qquad
        \odot
            (
                e^{i \alpha \l \vtil  }
                    \otimes
                        \ketbra {\util,\vtil} {0,0}[GG]
            )
                \\
    &=
        \intgdom d\util\, d\vtil\,
            (
                \id_L
                    \otimes
                        \ketbra {0,0} {\util,\vtil}[GG]
            )
                e^{-i \alpha \l \otimes \vGop}
                    \nonumber \\
        &\qquad \qquad
            \odot
                e^{i \alpha \l \otimes \vGop}
                    (
                        \id_L
                            \otimes
                                \ketbra {\util,\vtil} {0,0}[GG]
                    )
                        \\
    &=
        ( \id_L \otimes \ket {0,0}[G] )
            \Tr_G
                (
                    e^{-i \alpha \l \otimes \vGop}
                        \odot
                            e^{i \alpha \l \otimes \vGop}
                )
                    \nonumber \\
    &\qquad \qquad
        \times
            ( \id_L \otimes \bra {0,0}[G] )
                \\
    &=
        \Tr_G
            (
                e^{-i \alpha \l \otimes \vGop}
                    \odot
                        e^{i \alpha \l \otimes \vGop}
            )
            \otimes
                \ketbra {0,0}[GG]
    \,,
\end{align}
where in the third line, we used the fact that $\vtil \ket{\util, \vtil} = \vGop \ket{\util, \vtil}$ to promote the logical rotation to a logical-gauge controlled unitary; in the fourth line, we used
$
    \Tr_G
        =
            \int_\gdom d\util\, d\vtil\, \bra
                {\util,\vtil}[G]
                    \odot \ket {\util,\vtil}[G]
$;
and in the final line, we noted that $\Tr_G$ is a purely logical operator, so it commutes with the gauge pieces.

All told, the GKP error-correction channel~$\ecchan$ first applies a logical-gauge entangling operation~$e^{-i \alpha \l \otimes \vGop}$ and then takes the usual (MV SSD) gauge trace, \cref{eq:uvgaugetrace}, resulting in the logical state
\begin{align}
\label{eq:rhoLEC}
    \rhop_{L,\EC}[\rhop]
        &\coloneqq
            \Tr_G
                (
                    e^{-i \alpha \l \otimes \vGop}
                        \rhop
                            e^{i \alpha \l \otimes \vGop}
                )
\end{align}
as the logical qubit state, with the gauge mode always left in~$\ketbra{0,0}[GG]$. Thus,
\begin{align}
    \ecchan(\rhop)
        =
            \rhop_{L,\EC}[\rhop]
                \otimes
                    \ketbra{0,0}[GG]
    \,.
\end{align}
This channel is equivalent to performing error correction and averaging the resulting state over the possible syndrome outcomes.

An ordinary partial trace, such as $\Tr_G$, preceded by an entangling operation can be interpreted as a new partial trace with respect to a different tensor-product decomposition~\cite{zanardi2004quantum}. Thus, we may define
\begin{align}
\label{eq:trGEC}
    \Tr_{G_\EC}
        \coloneqq
            \Tr_G
                \circ
                    (
                        e^{-i \alpha \l \otimes \vGop}
                            \odot
                                e^{i \alpha \l \otimes \vGop}
                    )
\end{align}
as an inequivalent error-correction-based partial trace of a CV mode resulting in a different type of logical qubit, namely~$\rhop_{L,\EC}[\rhop] = \Tr_{G_\EC}(\rhop)$, as distinct from $\rhop_{L}[\rhop] = \Tr_G(\rhop)$. Recent work~\cite{shaw_2022} explores a tensor-product decomposition of a CV mode into a logical and gauge mode directly based on GKP error correction, and it remains an interesting open question to connect the present discussion to that framework.

\section{Summary and outlook}
\label{sec:connections}

We reviewed the Zak transform formalism in the context of continuous-variable quantum mechanics. We focused primarily on issues related to periodicity and quasi-periodicity of modular wavefunctions and Zak basis states that lie in a fundamental domain of area $2\pi$. By modifying the Zak transform to include additional periodicity, we introduced a stretched Zak basis spanned by basis kets that lie in a transformed domain whose area is not $2\pi$. Two consequences are that stretched-Zak kets obey different quasi-periodicity relations, and the stretched modular variables are no longer the same operators as Aharonov's, but squeezed versions thereof.

Modular wavefunctions are useful for working with the GKP code. One practical appeal is the compactness of the domain where they are defined. Ideal GKP states are represented in a simple form as ``two dots in the Zak patch'' (two-dimensional Dirac delta-distributions) whose amplitudes and phases specify the state. Good approximations to a GKP states have Zak transforms that are mainly distributed around these points---the narrower the distribution, the better the approximation. For these states, $\abs{\psi(u,v)}^2$ can be interpreted as the probability distribution that an ideal GKP state has been displaced in position and momentum by $u$ and $v$ respectively~\cite{glancy_error_2006,glancy_error_2006}.

Moreover, the Zak domain provides a pathway to more generally interpret the logical-qubit content of a CV state. By dividing the fundamental rectangular domain $\zakdom$ down the middle into two ``squares'', we associate the left square with state $\ket{0}$ state and the right square with state $\ket{1}$.%
\footnote{%
    The patches are only square when $\alpha=\rpi$, but it is convenient to
    always consider this parameter choice when representing the domain
    pictorially, as we have done in all figures.
}
We formalized this concept with a subsystem decomposition~\cite{pantaleoni_modular_2020} that decomposes the Zak basis into a tensor-product basis for a qubit and a gauge mode in the stretched Zak basis, $L^2(\zakdom) \cong \complex^2 \otimes L^2(\zakdom_G)$.

Decomposing quadrature displacements in the modular variable subsystem decomposition revealed that position shifts can introduce discrete, periodic logical operations, whereas momentum shifts generate continuous rotations of the logical-qubit subsystem. (Both of the displacements are accompanied by gauge-mode transformations.) This logical rotation may seem incompatible with GKP error correction because the latter can perfectly correct small shifts in either quadrature without introducing logical byproducts. This feature is indeed revealed in the SSD of GKP error correction, which includes a logical counter-rotation that undoes the rotation introduced by a momentum-shift error.

Finally, we embedded this corrective rotation in a new logical qubit---defined with respect to GKP error correction---that differs from the one found by taking the trace over the gauge-mode subsystem of the MV SSD. Recently, it was shown that it is possible to define a different subsystem decomposition also based on modular variables wherein small, correctable shifts in both position and momentum act purely as gauge-subsystem operations~\cite{shaw_2022}. In that case, GKP error correction acts exclusively as a measurement of the gauge mode (\emph{i.e.},~with no logical-correction component).

Of course, regardless of the SSD, the fundamental quantum operation is fixed---GKP error correction---so the resulting state obtained from it is the same. Since all SSDs provide complete bases, albeit with generally different tensor-product decompositions~\cite{zanardi2004quantum}, any of them can be used to represent a CV quantum operation.

The main difference arises in the representation of the information before GKP error correction occurs. The logical subsystem before GKP error correction in the case of Ref.~\cite{shaw_2022}, and also when using~$\Tr_{G_\EC}$ as defined in this work, but not when using $\Tr_G$ alone, is the same as the GKP-encoded logical state after error correction. Whether this fact has practical utility remains to be seen, but it is appealing in its direct encoding of the post-error-correction logical information into one of the two subsystems.

Looking forward, one might explore how the SSD in Ref.~\cite{shaw_2022} relates in detail to the one induced by~$\Tr_{G_\EC}$. It may be the case that the associated SSD bases differ at the gauge-mode level even though the logical state in both is equivalent (and based on GKP error correction). Finally, we note that possible extensions to this work may include developing SSDs based on generalizations of the Zak transform to different Hilbert spaces corresponding to generalizations of GKP states to different quantum systems, such as those discussed in Ref.~\cite{albert_robust_2020}.

\acknowledgments
We thank Mackenzie Shaw, Andrew Doherty, and Arne Grimsmo for useful discussions about Zak bases and subsystem decompositions. This work is supported by the Australian Research Council via the Centre of Excellence for Quantum Computation and Communication Technology (CQC$^2$T) (Project No.\ CE170100012) and the Centre of Excellence in Engineered Quantum Systems (EQUS) (Project No.\ CE170100009). GP~acknowledges further support by the US Army Research Office under Grant Number W911NF-21-1-0007. BQB~was additionally supported by the Japan Science and Technology Agency through the MEXT Quantum Leap Flagship Program (MEXT Q-LEAP).

\bibliographystyle{apsrev4-2_title}
\bibliography{references}

\end{document}